\documentclass[MSNbibl,nameyear,seceqn,dvips]{arxstspdf}
\usepackage{dcolumn}
\usepackage{graphicx}
\usepackage{flushend}
\usepackage{stfloats}

\setattribute{abstract}   {width}  {29pc}
\setattribute{keyword}    {width}  {29pc}

\volume{26}
\issue{3}
\pubyear{2011}
\firstpage{352}
\lastpage{368}
\doi{10.1214/11-STS364}

\makeatletter
\newcolumntype{d}[1]{D{.}{.}{#1}}
\makeatother

\begin{document}
\begin{frontmatter}

\title{A Problem in Particle Physics and Its Bayesian Analysis}
\runtitle{Particle Physics}

\begin{aug}
\author{\fnms{Joshua} \snm{Landon}\ead[label=e1]{jlandon@gwu.edu}},
\author{\fnms{Frank X.} \snm{Lee}\ead[label=e2]{fxlee@gwu.edu}}
\and
\author{\fnms{Nozer D.} \snm{Singpurwalla}\corref{}\ead[label=e3]{nozer@gwu.edu}}
\runauthor{J. Landon, F. X. Lee and N. D. Singpurwalla}

\affiliation{George Washington University}

\address{Joshua Landon is Assistant Professor, Department of Statistics, George Washington University,
Washington, District of Columbia 20052, USA \printead{e1}.
Frank X. Lee is Professor, Department of Physics, George Washington University,
Washington, District of Columbia 20052, USA \printead{e2}.
Nozer D. Singpurwalla is Professor, Department of Statistics,
George Washington University,
Washington, District of Columbia 20052,
USA \printead{e3}.}
\end{aug}

%
\begin{abstract}
There is a class of statistical problems that arises in several
\mbox{contexts}, the Lattice QCD problem of particle physics being one that has
attracted the most attention. In essence, the problem boils down to the
estimation of an infinite number of parameters from a finite number of
equations, each equation being an infinite sum of exponential
functions. By
introducing a latent parameter into the QCD system, we are able to identify
a pattern which tantamounts to reducing the system to a telescopic
series. A~statistical model is then endowed on the series, and inference about the
unknown parameters done via a Bayesian approach. A computationally intensive
Markov Chain Monte Carlo (MCMC) algorithm is invoked to implement the
approach. The algorithm shares some parallels with that used in the particle
Kalman filter. The approach is validated against simulated as well as data
generated by a~physics code pertaining to the quark masses of protons. The
value of our approach is that we are now able to answer questions that could
not be readily answered using some standard approaches in particle
physics.

The structure of the Lattice QCD equations is not unique to
physics. Such architectures also appear in mathematical biology, nuclear
magnetic imaging, network analysis, ultracentrifuge, and a host of other
relaxation and time decay phenomena. Thus, the methodology of this paper
should have an appeal that transcends the Lattice QCD scenario which
motivated~us.

The purpose of this paper is twofold. One is to draw attention
to a class of problems in statistical estimation that has a broad
appeal in
science and engineering. The second is to outline some essentials of
particle physics that give birth to the kind of problems considered here.
It is because of the latter that the first few sections of this paper are
devoted to an overview of particle physics, with the hope that more
statisticians will be inspired to work in one of the most fundamental areas
of scientific inquiry.
\end{abstract}

%

\end{frontmatter}

\section{Introduction and Overview}\label{sec1}

Lattice Quantum Chromodynamics, or Lattice\break QCD, is an actively researched
topic in particle phy\-sics. Many investigators in this field have received
the Physics Nobel Prize, the 2004 prize going to Gross, Politzer and
Wilczek, developers of the notion of ``asymptotic
freedom'' that characterizes\break QCD. Underlying the Lattice
QCD equations are issues of parameter estimation that have proved to be
challenging. Essentially, one needs to estimate an infinite number of
parameters from a finite number of equations, each equation being an
infinite sum of exponential functions.

The approach proposed here is Bayesian; it is dri\-ven by a computationally
intensive Markov Chain Monte Carlo (MCMC) implementation. However, to invoke
this approach, we need to introduce a \mbox{latent} parameter and then explore the
``anatomy'' of the QCD equations. This
reveals a pattern, which when harnessed with some reasonable statistical
assumptions provided a pathway to a solution. The inferen\-ces provided
by our
approach were successfully vali\-dated against simulated as well as real data.
\mbox{However}, the real value of our approach is that it is able~to~ans\-wer
questions that could not be answered using some of the conventional
approaches of particle physics. The approach can therefore be seen as an
addition to the lattice field theorists' data analysis tool
kit.

The structure of the Lattice QCD equations is not as specialized as one is
inclined to suppose. Indeed, such equations also appear in other
contexts of
engineering, physics, nuclear magnetic imaging and mathematical biology
where they go under the label of ``exponential
peeling;'' see Section \ref{sec3.1}. Our focus on the physics
scenario is due to the fact that this is how we got exposed to the general
problem addressed here.

This paper is directed toward both statisticians and physicists, and could
serve as an example of the interplay between the two disciplines. The former
may gain an added appreciation of problems in modern physics that can be
addressed via statistical methods. In the sequel, they may also get to know
more about particle physics and the beautiful theories about it that Mother
Nature has revealed. It is, with the above in mind, that Section \ref{sec2} is devoted
to an overview of aspects of particle physics, its associated terminology
and the awe inspiring discoveries about it. Reciprocally, the
physicists may
benefit by exposure to some modern statistical technologies that can be
brought to bear for addressing problems that may have caused them
some\vadjust{\goodbreak}
consternation.

Section \ref{sec2} gives an overview of some essentials of particle physics, and the
ensuing Lattice QCD equations. This section, written by a nonphysicist
(NDS) but reviewed by a physicist (FXL), has been developed by fusing
material from a variety of sources, some notable ones being Pagels (\citeyear{Pag82}),
Dzierba,\break Meyer and Swanson (\citeyear{AmericanScientist00}), Yam (\citeyear{ScientificAmerican93}),
\citet{ScientificAmerican06} and Frank
Wilczek's (\citeyear{Wil05}) Nobel lecture. Interjected throughout this section are a
few comments of historical interest; their purpose is to inform a
nonphysicist reader about the individuals who have contributed to the
building of a magnificent edifice. Section \ref{sec2} concludes with a graphical
display of the structure of matter via a template that is familiar to
statisticians, in particular, those working in network theory and in
reliability.

Section \ref{sec3} pertains to an anatomy of the Lattice QCD equations and the
resulting mathematical pattern that it spawns. It is not necessary to read
Section \ref{sec2} (save perhaps for an inspection of Figure \ref{fig2.5}) in order to read
Section \ref{sec3}, which is where this paper really begins; indeed, Section \ref{sec2} could
have been delegated to an Appendix. Section \ref{sec3} is a foundation for the rest
of the paper. It is here that the inferential problem is introduced along
with its accompanying notation and terminology. Section \ref{sec3.1} gives a broad
overview of the several other scenarios in science and engineering
where the
Lattice QCD type equations also arise. Of particular note are the several
examples in mathematical biology wherein the QCD like equations are often
discussed.

Section \ref{sec4} pertains to the statistical model that~the material of
Section \ref{sec3}
creates, and an outline of the MCMC approach that is used to estimate the
para\-meters of the model. These are the parameters that are of interest to
physicists and other scientists. Section~\ref{sec5} pertains to validation against
simulated and actual data and proof of principles. Section~\ref{sec6} pertains to
some suggestions for extending the work~do\-ne here, and strategies for
overcoming some of~the~en\-countered difficulties. Section \ref{sec7} concludes the~paper.

\begin{figure}

\includegraphics{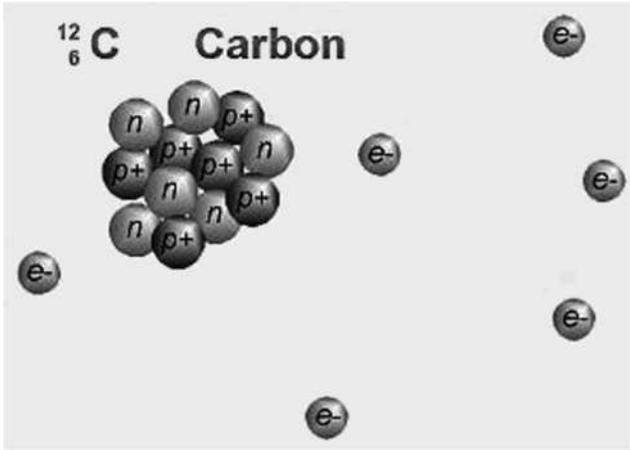}

  \caption{Architecture of a carbon atom.}\label{fig2.1}
\end{figure}

\begin{figure*}[b]\vspace*{-3pt}

\includegraphics{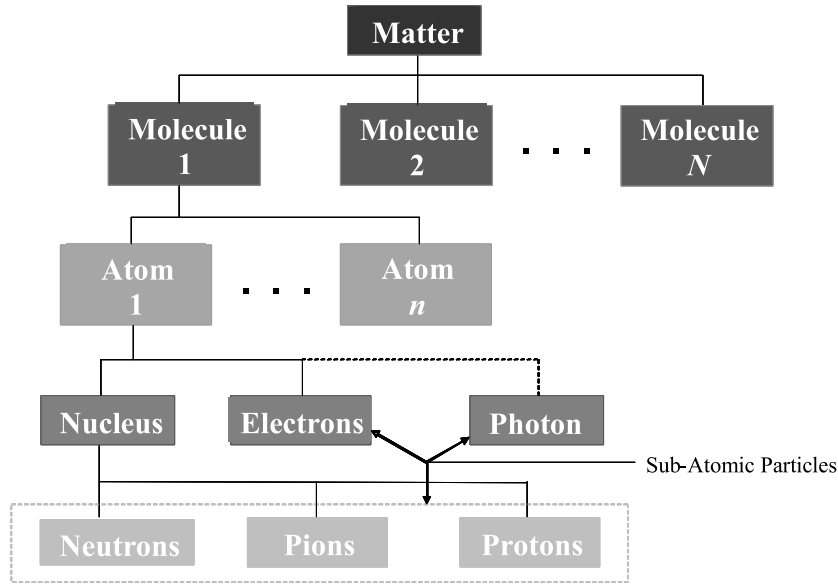}

  \caption{The structure of matter (circa 1946).}\label{fig2.2}
\end{figure*}

Since the Lattice QCD equations can be seen as a prototype for similar
equations that arise in other scientific endeavors, this paper also serves
as an invitation to other statisticians to develop approaches for solving
such equations using methods more sophisticated and/or alternate to the one
we have entertained.

\vspace*{-3pt}\section{Essentials of Particle Physics}\vspace*{-3pt}\label{sec2}

The smallest quantity of anything we can see or feel is a \textit
{{molecule}}, and all matter is made up of~mo\-lecules,\vadjust{\goodbreak} which in turn are made
up of {\textit{atoms}}. Molecu\-les and atoms are called {\textit{particles}},
and the physics that describes the interactions between
the particles is known as {\textit{particle physics}}; see, for
example, Griffiths (\citeyear{Gri87}).

An atom consists of {\textit{electrons}}, which carry a negative
charge, and the electrons are centered around a~{\textit{nucleus}}
that is made up of {\textit{protons}} that carry a~po\-sitive charge,
and {\textit{neutrons}} that carry no charge. Fi\-gure \ref{fig2.1} illustrates
the architecture of a carbon atom which has six electrons, six protons and
six neutrons; it is denoted ${}_{6}^{12}$C.

The protons and the neutrons are held together within the nucleus by a
nuclear glue called the {\textit{pion}}. Similarly, the protons and
the electrons are held together within the atom by a glue called the
{\textit{photon}}. The pions are said to be carriers\vadjust{\goodbreak} (or mediators) of the
{\textit{strong force}} (or the {\textit{nuclear
force}}), and
the photons are carriers of the {\textit{electromagnetic force}}.
Physicists look at the nuclear glues as force carrying particles, and thus
collectively regard the electrons, the neutrons, the photons, the pions and
the protons as {\textit{subatomic particles}}. Figure \ref{fig2.2} displays
the structure of matter as understood around the 1946 time frame. The dotted
lines of Figure~\ref{fig2.2} indicate the glued members.

\begin{figure*}

\includegraphics{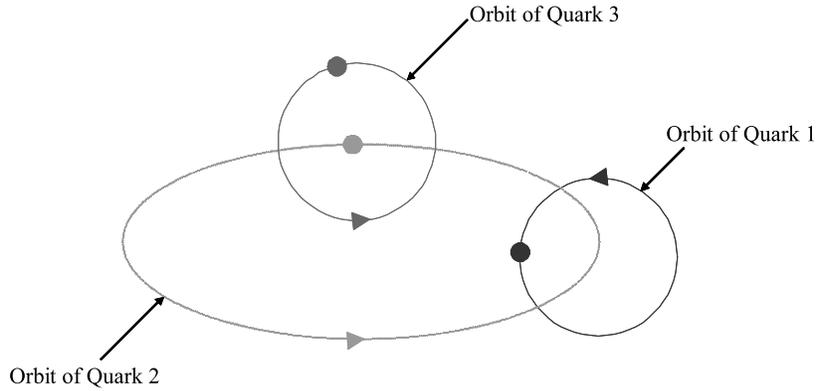}

  \caption{Illustration of a quark orbit.}\label{fig2.3}
  \vspace*{-3pt}
\end{figure*}

\begin{figure*}[b]
\vspace*{-3pt}
\includegraphics{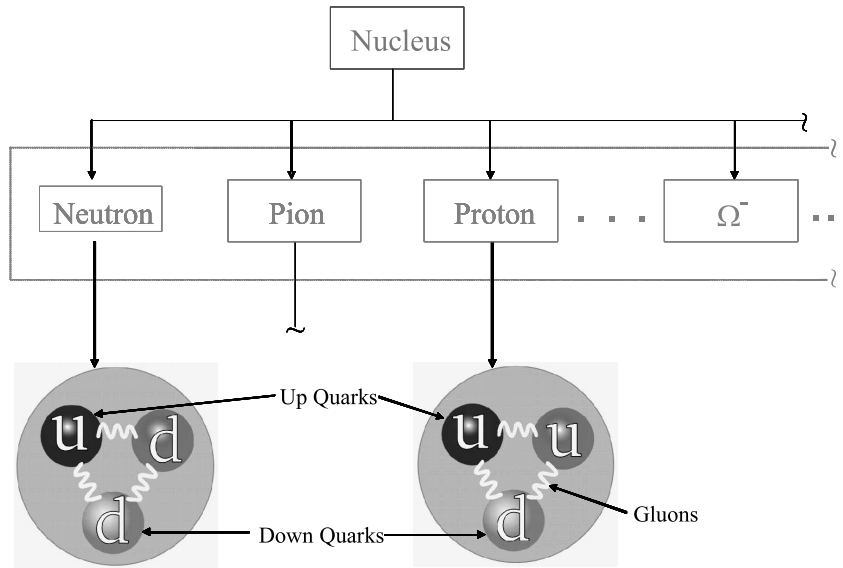}

  \caption{The quark structure of hadrons.}\label{fig2.4}
\end{figure*}

In 1911, when Rutherford announced the structure of the atom, the existence
of electrons and protons was known. The neutron, as a major constituent of
the nucleus, was discovered in 1932 by Chadwick, and the pion was discovered
in 1946. But these discoveries were just the tip of the iceberg. Many more
subatomic particles have subsequently been discovered. Collectively, these
subatomic particles are now called {\textit{hadrons}}. Physicists
speculate that there exist an infinite number of such hadrons. This
discovery of hadrons was made possible by {\textit{accelerators}},
which are essentially microscopes for matter.

The invention of the accelerators opened up the subnuclear world with the
experimental discovery of thousands of new particles. The question thus
arose as to what the hadrons could be saying about the ultimate
structure of
matter.

\subsection{The Quark Structure of Matter}\label{sec2.1}

The current view is that hadrons are composite objects made out of more
fundamental particles cal\-led {\textit{quarks}}, and no one has ever
seen a quark! This point of view came about in the early 1960s when Murray
Gell-Mann discovered that the hadrons organized themselves into classes (or
families) based on a mathematical symmetry. An easy way to understand why
this organizational principle worked is to assume that the hadrons are made
up of quarks, only three of which were needed to build the hadrons. These
quarks were named the {\textit{up quark}}, the {\textit{down
quark}} and the {\textit{strange quark}}. For example, a proton has
two up quarks and one down quark, whereas a neutron has two down quarks and
one up quark. In general, every hadron is made up of quarks that orbit
around each other in a specific configuration, each configuration resulting
in a hadron. Figure \ref{fig2.3} is an illustration of a quark orbit.

\begin{figure*}

\includegraphics{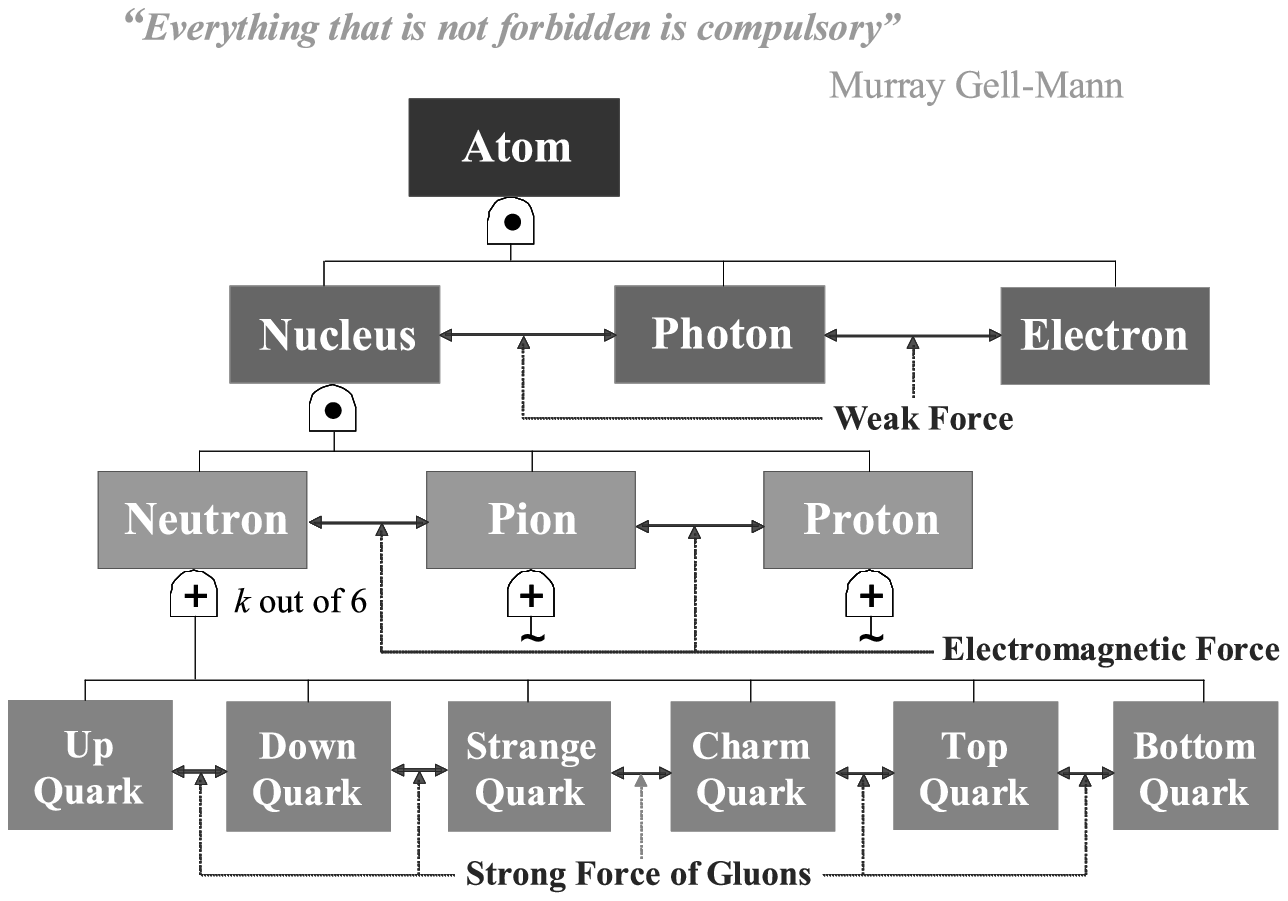}

  \caption{Matter as a coherent system.}\label{fig2.5}
  \vspace*{-9pt}
\end{figure*}

Since there could be several orbit configurations, there ought to be an
infinite number of hadrons.\vadjust{\goodbreak} The essence of Gell-Mann's idea is that
hadrons
are bound states of quarks, just like how the atoms are bound states of
electrons, neutrons and protons. Furthermore, Gell-Mann postulated that
there ought to exist a force carrying particle, called the
{\textit{gluon}}, that holds the quarks together. The gluon is said to be the carrier
of the {\textit{strong force}}. Figure \ref{fig2.4} illustrates the quark
structure of a hadron.

The quark model was purely a theoretical cons\-truct. Its validity was
affirmed when Gell-Mann~used it to postulate in 1962 the existence of a
particle never seen before. This was a scientific breakthrough of the
highest order! It showed that discoveries in physics can come from
mathematical patterns---not just the laboratory. For unraveling the
mathematical symmetries of the hadron, Gell-Mann received the 1969 Nobel
Prize in Physics.

Figure \ref{fig2.5} gives a pictorial representation of the quark structure of matter
using a template that is familiar to statisticians. It represents an
atom as a~coherent (or logical) system with quarks as the basic building blocks of
the system. The logic symbols of ``and''
and ``or'' are represented by 
and respectively. The neutrons and
the protons can be~re\-garded as subsystems, and the gluons, photons and the
pions that link the quarks, the nucleus and the electrons can be seen as
the structure (or link) functions of the system (cf. Barlow and
Proschan,
\citeyear{BarPro75}). These are the carriers of the strong force and the electromagnetic
force, respectively. Figure~\ref{fig2.5} contains Gell-Mann's famous quote that
``everything that is not forbidden is compulsory;''  the logical systems analogue to this
quote is the notion of ``irrelevance.''

\vspace*{-2pt}\subsection{Quantum Chromodynamics and Lattice QCD}\vspace*{-2pt}\label{sec2.2}

The theory of QCD can be thought of as a recipe for producing hadrons from
quarks and gluons. Since quarks and gluons make up most of the known mass
of the physical world, unraveling the quark structure of matter is the key
to an understanding of the physical world, and thus the importance of the
subject of this paper.

The QCD theory was successful in enunciating~the properties of the hadrons.
However, its \mbox{complexity} made its use for predicting unobservable quantum
quantities, like quark masses,\vadjust{\goodbreak} almost impossible. This is because solving
the QCD equation (which is just one line) by analytical methods is
difficult. The current approach is to solve the QCD equation
numerically, by
discretizing it over a space--time lattice. Lattice QCD refers to the
representation of space--time as a scaffold in four dimensions wherein the
quarks rest on the connecting sites, and the gluons as connections between
the lattice points.

The scaffold is first restricted to a finite volume; it is then replicated
with periodic boundary conditions. All this entails on the order of 100
million billion arithmetic operations on typical lattices; this is one
example as to why physicists need supercomputers. Lattice QCD has been able
to explain as to why a free quark has not been seen and will not be seen;
this is because it will take an infinite amount of energy to isolate a quark.

Lattice QCD, being an approximation to the QCD, improves as the lattice
points increase indefinitely and as the volume of the lattice grid expands.
In so doing it opens up avenues for statistical methods to enter the
picture. Physicists have explored some of these avenues, one of which
is the
focus of this paper; see Section \ref{sec3} below.

\section{The Underlying Problem: QCD equations}\label{sec3}

With Lattice QCD, an archetypal scenario is the estimation of an infinite\vadjust{\goodbreak}
number of parameters from a finite number of equations. The left-hand side
of each equation is the result of a physics based Monte Carlo run, each
run taking a long time to complete. Thus, there are only a finite
number of
runs. For example, a {\textit{meson correlator}}, $G(t|\cdot)$,
takes the form (cf. Lepage et al., \citeyear{Lepetal02})
%
\begin{equation}\label{eq3.1}
\qquad G(t|\cdot)=\sum_{n=1}^{\infty}A_{n}e^{-E_{n}t}\quad \mbox{for }t=0,1,2,\ldots,
\end{equation}
where the parameters $A_{n}$ denote the {\textit{amplitude}},
and $E_{n}$ denote the {\textit{energy}}. Also, $E_{1}\,{<}\,E_{2}\,{<}\cdots
\,{<}\,E_{n}\,{<}\cdots$.

Interest centers around the estimation of $A_{n}$\break and~$E_{n}$,
$n=1,2,\ldots,$ based on $G(t|\cdot)$, estimated as $\widehat{G}(t|\cdot), t=0,1,\ldots,k$,
for some finite $k$ [23 in the case of Lepage et al.
(\citeyear{Lepetal02})]. The physics codes which generate the $\widehat{G}(t|\cdot
)$'s do
not involve the $A_{n}$'s and the~$E_{n}$'s, and are autocorrelated, thus
the label ``correlator.''  The
physics codes
also provide estimates of the autocorrelation matrix.

Deterministic approaches to solve for the $A_{n}$'s and the $E_{n}$'s cannot
be invoked, and statistical approa\-ches involving curve fitting by
chi-square, maximum likelihood and empirical Bayes have proved to be~un\-satisfactory
(cf. Morningstar, \citeyear{Mor02}). For an apprecia\-tion of these
efforts, see Lepage et al. (\citeyear{Lepetal02}), Fiebig (\citeyear{Fie}) and Chen et al. (\citeyear{Cheetal});
the latter authors propose what they call a ``sequential
empirical Bayes approach.''  However, empirical Bayes
approaches use observed data to influence the choice of priors, and, as
asserted by Morningstar (\citeyear{Mor02}), are a~violation of the Bayesian philosophy.
Indeed, Fiebig (\citeyear{Fie}) states that ``Bayesian inference has~too long been \mbox{ignored} by
the lattice community as~an analysis\break tool.
\ldots The method should be given~serious consideration as an alternative for
conventional ways.''

Bayesian approaches alternate to ours have been considered by Nakahara, Asakawa and Hatsuda\break
(\citeyear{NakAsaHat99}). These authors entertain the use of maximum entropy priors, but,
as claimed by Lepage et~al. (\citeyear{Lepetal02}), the accuracy of their estimator of $E_{2}$
is inferior to those obtained using other approaches. Because priors
based on the principle of maximum entropy result in default priors, such
priors also violate the Bayesian philosophy. The approach of Lepage et al.
(\citeyear{Lepetal02}) is Bayesian in the sense that prior information is used to
augment a
chi-square statistic which is then minimized. We find this work valuable
because it articulates the underlying issues and provides a framework for
examining the anatomy of the QCD equations, which enables us to
identify a
pattern, which in turn enables us to invoke the Bayesian approach we\vadjust{\goodbreak} propose.

\subsection{Relevance to Other Scenarios in Science and Engineering}\label{sec3.1}

The Lattice QCD architecture of equation (\ref{eq3.1}) is not unique to physics.
They occur in several other scenarios in the physical, the chemical, the
engineering and the biological sciences, a few of which are highlighted
below. Most attempts at estimation of the underlying parameters have
involved least squares or numerical techniques based on local linearization
with iterative improvements. Besides lacking a theoretical foundation
vis-%
\`{a}-vis the requirement of coherence (cf. Bernardo and Smith, \citeyear{BerSmi94},
page 23), techniques have proved notoriously unreliable and not robust to
slight changes in the experimental data (cf. Hildebrand, \citeyear{Hil56}).

\vspace*{2pt}\subsubsection*{Mathematical biology: exponential peeling in
compartment systems}

When considering radioactive tra\-cers used for studying transfer rate of
substances in living systems (cf. Robertson, \citeyear{Rob57}; Rubinow, \citeyear{Rub75},
page~125),
sums of exponentials are encountered. Here, the $G(t|\cdot)$ of equation
(\ref{eq3.1}) represents the concentration of a substance, the $t$'s are integer
values of time, and the $A_{i}$'s and the $E_{i}$'s are constants that
need to
be estimated. Here interest generally centers around the case of $n=2$, and
the coefficients~$A_{n}$ and~$E_{n}$ of equation (\ref{eq3.1}) are negative. An ad
hoc graphical procedure called \textit{the method of exponential
peeling} is
used to estimate the parameters (cf. Smith and Morales, \citeyear{SmiMor44}, Perl,
\citeyear{Per60};
van Liew, \citeyear{Van67}).

Some other scenarios in biology where the Lattice QCD type equations appear
are in bone metabolism studies and cerebral blood flow (cf. Glass and de
Garreta, \citeyear{GlaGer67}), and in biological decay (cf. Foss, \citeyear{Fos69}). In the latter
context, Dyson and Isenberg (\citeyear{DysIse71}) consider for fluorescence decay an
equation of the type%
\[
y(t)=\sum_{j=1}^{m}\alpha_{j}\exp(-t/\tau_{j}),\quad 0\leq
t\leq T,
\]
where $y(t)$ represents ``moments of the
fluorescence,'' $\alpha_{j}$'s the amplitudes [the
$A_{n}$%
's of equation (\ref{eq3.1})], and the $\tau_{j}$'s are time constants
corresponding to the~$E_{n}$'s of equation (\ref{eq3.1}). Here the $\alpha_{j}$'s are zero for
$j\geq m+1$.

\subsubsection*{Gene expression data}

When considering a time series of gene expression data (cf. Giurcaneanu et~al.,
\citeyear{GiuTabAst05}), a system of equations paralleling that of equation (\ref{eq3.1}) arises
again. In this context $G(t|\cdot)$ represents ``mRNA
concentrations'' as a function of time, and the
parameters $A_{n}$ and $E_{n}$ describe interactions between the genes. In the\vadjust{\goodbreak}
gene expression context, as in the Lattice QCD context, the parameters $E_{n}$ are increasing in~$n$.

\vspace*{-2pt}\subsubsection*{Nuclear magnetic resonance (NMR)}

NMR experiments often generate data that are modeled as the sum of
exponentials (cf. Bretthorst et al., \citeyear{Breetal05}). Experiments relying on
NMR to
probe reaction kineticis, diffusion, molecular dynamics and xenobiotic
metabolism are some of the applications where parameter estimates provide
insight into chemical and biological processes. See, for example, Paluszny
et al. (\citeyear{Paletal08}) who study brain tissue segmentation from NMR data.

Here one considers equations of the type%
\[
d_{i}=C+\sum_{j=1}^{m}A_{j}\exp\{-\alpha_{j}t_{i}\}+n_{i},
\]
where $m$ is the number of exponentials and $d_{i}$ a data value
sampled at $%
t_{i}$. The parameters of interest are the decay rate constants $\alpha
_{j}$%
, the amplitudes $A_{j}$ and the constant offset $C$. The $n_{s}$'s are the
error terms.

\vspace*{-2pt}\subsubsection*{Electromechanical oscillations in power
systems}

Equations entailing the sum of exponentials are also encountered in the
context of low frequency electro\-mechanical oscillations of interconnected
power systems, the impulse response of linear systems in networks,
ultracentrifuge and a host of other relaxation and time-decay phenomena (cf.
Dyson and Isenberg, \citeyear{DysIse71}). For example, in the electromagnetic oscillations
scenario, Sanchez-Gasca and Chow (\citeyear{SanCho99}) encounter an equation analogous to
our equation~(\ref{eq3.1}) with $G(t|\cdot)$ denoting a signal and $A_{n}$
connoting a~signal residue associated with the ``mode'' $E_{n}$.

To summarize, the relationships of the type given by equation (\ref{eq3.1})
arise in
so many contexts of science and engineering that it seems to be
quintessential, and almost some kind of law of nature. The Lattice QCD
problem considered here can therefore be seen as a prototype and a
convenient platform to exposit a statistical problem of general
applicability. In most of the application scenarios described above,
statistical methods have been used, many ad hoc, some empirical Bayesian
and a few Bayesian (under the rubric of maximum entropy). Many of these
methods have not exploited an underlying telescopic pattern in these
equations which makes an appearance when a latent parameter is introduced
into the system, and inference about the latent parameter
made.

\vspace*{-2pt}\subsection{Anatomy of the Lattice QCD equations}\vspace*{-2pt}\label{sec3.2}

An examination of equation (\ref{eq3.1}) yields the following boundary\vadjust{\goodbreak}
conditions. $G(0|\cdot)=\sum_{n=1}^{\infty}A_{n}$, implying that the $A_{n}$'s are
constrained. When $t\rightarrow\infty$, $G(t|\cdot)\,{=}\,0$, which implies
that for large values of $t, A_{n}$ and the $E_{n}$ cannot be individually
estimated. Thus, simulating $G(t|\cdot)$ for large $t$ does not have a~payback;
consequently, it is futile to do such a~simulation.

Since the $E_{n}$'s increase with $n$, we may, as a start, reparameterize
the $E_{n}$'s as $E_{n}-E_{n-1}=c$, for some \textit{unknown} $c$, $c>0$,
for $n=2,3,\ldots.$ It will be argued later, in Section \ref{sec6.1}, that $c$
is a
latent parameter. Thus,
%
\begin{equation}\label{eq3.2}
E_{n}=E_{1}+(n-1)c,\quad n=2,3,\ldots
\end{equation}
with $E_{1}$ and $c$ unknown. With the above assumption in place, a
parsimonious version of the Lattice QCD equation takes the form%
%
\begin{eqnarray}\label{eq3.3}
G(t|\cdot)=e^{-E_{1}t}\sum_{n=1}^{\infty}A_{n}e^{-(n-1)ct},\\
\eqntext{
t=0,1,2,\ldots.}
\end{eqnarray}
With $c$ fixed, the parsimonious model given above reveals the following
features:

\begin{longlist}[(a)]
\item[(a)] When $t$ is \textit{small}, the number of $A_{n}$'s entering
equation (\ref{eq3.3}) is large; indeed, infinite when $t=0$.

\item[(b)] When $t$ is \textit{large}, the number of $A_{n}$'s we need
to consider is small, because the combination of a~large~$t$ with any $n$
will make the term $A_{n}\exp(-(n-1)ct)$ get small enough to be ignored.

\item[(c)] Moderate values of $t$ and $n$ will also make the above term
small, causing $A_{n}$ to be irrelevant.\vspace*{-3pt}
\end{longlist}

\begin{figure}
\includegraphics{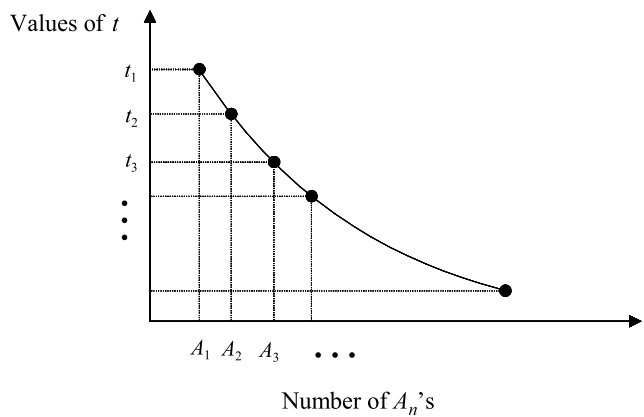}
\vspace*{-3pt}
  \caption{Number of $A_{n}$'s as a function of $t$.}\label{fig3.1}
\vspace*{-5pt}
\end{figure}

Figure \ref{fig3.1} illustrates the feature that as $t$ gets large, the number
of $A_{n}$'s one needs to consider gets small.

As a consequence of the above, for any fixed $c$, we can find a $t_{1}$ such
that in the expression%
\begin{eqnarray*}
&&e^{-tE_{1}}\bigl[ A_{1}+A_{2}e^{-ct}+A_{3}e^{-2ct}+\cdots
\\[-3pt]
&&\hspace*{58pt}\quad{}+A_{n}e^{-(n-1)ct}+\cdots\bigr] ,
\end{eqnarray*}
all the terms, save for $A_{1}$, are essentially\vadjust{\goodbreak} zero.

Similarly, we can find a $t_{2}$, $t_{2}<t_{1}$, such that all the terms
save for $A_{1}$ and $A_{2}e^{-ct_{2}}$ get annihilated. Continuing in this
vein, there exists a sequence $t_{k}<t_{k-1}<\cdots<t_{2}<t_{1}$, such that
all that matters are the terms associated with $A_{1},A_{2},\ldots,A_{k}$.
In what follows, we suppose that $k$ is \textit{specified}.

Thus, for any fixed $c$ and $k$, with $t_{1}>$ $t_{2}>\cdots>t_{k}$ chosen
in the manner described above, our parsimonious version of the Lattice QCD
equations \textit{telescope} as follows:%
%
\begin{eqnarray}\label{eq3.4}
G(t_{1}|\cdot) &=&e^{-E_{1}t_{1}}A_{1}, \nonumber\\
G(t_{2}|\cdot) &=&e^{-E_{1}t_{2}}(A_{1}+A_{2}e^{-ct_{2}}),
\nonumber\\
\qquad G(t_{3}|\cdot)
&=&e^{-E_{1}t_{3}}(A_{1}+A_{2}e^{-ct_{3}}+A_{3}e^{-2ct_{3}}),
\nonumber
\\[-8pt]
\\[-8pt]
\nonumber
&\vdots& \\
G(t_{k}|\cdot)
&=&e^{-E_{1}t_{n}}\bigl(A_{1}+A_{2}e^{-ct_{k}}+\cdots\nonumber\\
&&\hspace*{48pt}{}+A_{k}e^{-(k-1)ct_{k}}\bigr).\nonumber
\end{eqnarray}
To summarize, by introducing the constant $c$, fixing a $k$, and identifying
an underlying pattern in the Lattice QCD equations, we have reduced the
problem to the case of $k$ equations and $(k+2)$ unknowns, $A_{1},\ldots
,A_{k},E_{1}$ and $c$. The choice of what $k$ to choose is determined
by the
number of physics code based estimates $\widehat{G}(t), t=0,1,\ldots,k$,
that can be done and are available.

\section{Statistical Model: Solving the QCD Equations}\label{sec4}

Many have expressed the view that it would be considered good progress if
trustworthy estimates of just $A_{1},A_{2},E_{1}$ and $E_{2}$ can be had.
The other pairs $(A_{3},E_{3})$, $(A_{4},E_{4}),\ldots,$ can be considered
later; see Section \ref{sec6}. Thus, we start by focusing attention on the first two
equalities of equation (\ref{eq3.4}); that is, the case $k=2$ and some fixed $c$.
Specifically, we consider%
%
\begin{eqnarray}\label{eq4.1}
G(t_{1}|A_{1},E_{1}) &=&e^{-E_{1}t_{1}}A_{1}\quad \mbox{and}
\nonumber
\\[-8pt]
\\[-8pt]
\nonumber
\qquad G(t_{2}|A_{1},E_{1},A_{2},c) &=&e^{-E_{1}t_{2}}(A_{1}+A_{2}e^{-ct_{2}}).
\end{eqnarray}
If $y_{i}\,{=}\,$ $\widehat{G}(t_{i}|\cdot), i\,{=}\,1,2,$ denotes the physics code
based evaluations of $G(t_{i}|\cdot)$, then our aim is to estimate~$A_{1}$,
$E_{1}, A_{2}$ and $c$, in light of $y_{1}$ and $y_{2}$. To set up
our~like\-lihoods, we take a lead from what has been done by Nakahara, Asakawa and Hatsuda
(\citeyear{NakAsaHat99}), and by Lepage et al. (\citeyear{Lepetal02}), to write%
%
\begin{eqnarray}\label{eq4.2}
Y_{1} &=&G(t_{1}|\cdot)+\varepsilon_{1}\quad\mbox{and}
\nonumber
\\[-8pt]
\\[-8pt]
\nonumber
Y_{2} &=&G(t_{2}|\cdot)+\varepsilon_{2},
\end{eqnarray}
where $\varepsilon_{i}\sim N(0,\sigma_{i}^{2}), i=1,2$, and\vadjust{\goodbreak}
$\operatorname{Corr}(\varepsilon
_{1},\varepsilon_{2})=\rho_{12}$.

Besides providing $y_{1}$ and $y_{2}$, the physics codes also provide
$\sigma_{1}^{2}$, $\sigma_{2}^{2}$ and $\rho_{12}$. As a consequence, the
statistical model boils down to the bivariate normal distribution,%
%
\begin{eqnarray}\label{eq4.3}
\left[
\matrix{
Y_{1} \cr
Y_{2}%
}
\right]
\sim N\left( \left[
\matrix{
G(t_{1}) \cr
G(t_{2})%
}
\right] ,\left[
\matrix{
\sigma_{1}^{2} & \rho_{12}\sigma_{1}\sigma_{2} \cr
\rho_{12}\sigma_{1}\sigma_{2} & \sigma_{2}^{2}}
\right] \right) .\hspace*{-40pt}
\end{eqnarray}
Writing out a likelihood function for the unknowns $A_{1}, E_{1}, A_{2}$
and $c$, based on equation (\ref{eq4.3}), is a~straightforward matter. However, we
need to bear in mind that since the parameters $A_{1}$ and $E_{1}$
appear in
both $G(t_{1}|\cdot)$ and $G(t_{2}|\cdot)$, both $y_{1}$ and $y_{2}$
provide information about $A_{1}$ and $E_{1}$, with $y_{2}$ providing
information about $A_{2}$ and $c$ as well. To exploit this feature, we
construct our likelihoods based on the marginal distribution of
$Y_{1}$, and
the conditional distribution of~$Y_{2}$ given $Y_{1}$. That is, on%
%
\begin{equation}\label{eq4.4}
Y_{1}\sim N( A_{1}e^{-E_{1}t_{1}},\sigma_{1}^{2})
\end{equation}
and%
%
\begin{eqnarray}\label{eq4.5}
&&( Y_{2}|Y_{1}\,{=}\,y_{1}) \!\sim\! N\biggl(\!G(t_{2}|\cdot)\,{+}\,\rho_{12}%
\frac{\sigma_{2}}{\sigma_{1}}\bigl(y_{1}\,{-}\,G(t_{1}|\cdot)\bigr),\hspace*{-30pt}
\nonumber
\\[-8pt]
\\[-8pt]
\nonumber
&&\hspace*{160pt}\sigma_{2}^{2}(1\,{-}\,\rho_{12}^{2})\!\biggr).\hspace*{-30pt}
\end{eqnarray}
Specifically, the likelihood of $A_{1}$ and $E_{1}$, with $y_{1}$~fi\-xed, is%
%
\begin{equation}\label{eq4.6}
\qquad\mathcal{L}(A_{1},E_{1};y_{1})\propto\exp\biggl[ -\frac{(
y_{1}-A_{1}e^{-E_{1}t_{1}}) ^{2}}{2\sigma_{1}^{2}}\biggr] ,
\end{equation}
and the likelihood of $A_{1}$, $E_{1}$, $A_{2}$ and $c$, with $y_{2}$ fixed,
and the effect of $y_{1}$ incorporated via the posterior distribution
of $%
A_{1}$ and $E_{1}$, is of the form
%
\begin{eqnarray}\label{eq4.7}
&&\mathcal{L}(A_{1},E_{1},A_{2},c;y_{1},y_{2})\nonumber\\
&&\quad\propto\exp \biggl[-\biggl\{ y_{2}-\bigl(
e^{-E_{1}t_{2}}(A_{1}+A_{2}e^{-ct_{2}}) \bigr)
\nonumber
\\[-8pt]
\\[-8pt]
\nonumber
&&\hspace*{49pt}\qquad{}+\rho_{12}\frac{\sigma
_{2}}{%
\sigma_{1}}( y_{1}-A_{1}e^{-E_{1}t_{1}}) \biggr\}
^{2}\\
&&\hspace*{88pt}\qquad{}\cdot[2\sigma
_{2}^{2}( 1-\rho_{12}^{2}) ]^{-1}\biggr] .\nonumber
\end{eqnarray}
In the above development, the covariance matrix is provided by the physics
code. As suggested by a~referee, a deeper investigation of this matrix may
be called for, because with increasing $t$, the variances are likely to
increase, posing computational challenges to the proposed approach.

\subsection{Specification of the Prior Distributions}\label{sec4.1}

To implement our Bayesian approach, we need to make assumptions about
conditional independence,\vadjust{\goodbreak} and assign prior distributions for the unknown
parameters. The priors that we end up choosing in Section \ref{sec5} are not
based on
knowledge of the underlying physics, but are proper priors based on an
appreciation of the material in Morningstar (\citeyear{Mor02}), Lepage et al. (\citeyear{Lepetal02})
and Fleming (\citeyear{Fl05}).

The $A_{i}$'s are supposedly between 0 and 1, and no relationship between
them has been claimed. Thus, it is natural to assume that $A_{1}$ and $A_{2}$
are apriori independent, and have a beta distribution on $(0,1)$ with
parameters $(\alpha,\beta)$; we denote this as $\mathcal
{B}(A_{i};\alpha
,\beta)$, $i=1,2$. The relationship between $E_{1}$ and $c$ is less
straightforward. We conjecture that the larger the~$E_{1}$, the smaller
the $%
c$, and that $E_{1}$ can take values over $(0,\infty)$. It is
therefore  reasonable to assume that the prior on $E_{1}$ is a gamma distribution with
scale parameter $\eta$ and shape parameter $\lambda$; we denote this
by $\mathcal{G}(E_{1};\eta,\lambda)$. Some other meaningful choices for
a~prior on $E_{1}$ could be a Weibull, or a Pareto, the~lat\-ter being
noteworthy as a fat-tailed distribution. To encapsulate the dependence
between $E_{1}$ and $c$, we suppose that, given $E_{1}$, $c$ has a uniform
distribution over $(0,\omega/E_{1})$, for some $\omega>0$. Finally, we
also assume that $E_{1}$ and $c$ are independent of all the $A_{i}$'s.

The above choice of priors, with user specified hyperparameters $\alpha
$, $%
\beta$, $\omega$, $\lambda$ and $\eta$, is illustrative. In principle,
any collection of meaningful priors can be used, since the ensuing inference
is done numerically via a Markov chain Monte Carlo (MCMC) approach.

Lepage et al. (\citeyear{Lepetal02}), and also Morningstar (\citeyear{Mor02}), seem to use independent
Gaussian priors for the parameters in question---see equations (8) and (11)
respectively. Indeed, Morningstar (\citeyear{Mor02}) makes the claim that
``practitioners often restrict the choice of a prior to some
familiar distributional form.''  The restricted parameter
space makes the choice of Gaussian priors questionable. An overview of how
the MCMC is invoked here is given next.

\subsection{An Outline of the MCMC Excercise}\label{sec4.2}

The telescopic nature of the Lattice QCD equations suggests that the MCMC
will have to be conducted in the following three phases:

\textit{Phase} I. Using $E_{1}^{(0)}$ as a
starting value and $y_{1}$ as data, obtain the posterior distribution
of $A_{1}$ and~$E_{1}$ via equation (\ref{eq4.6}) as the likelihood, and 1,000 iterations
of the MCMC run.

\textit{Phase} II. Using $c^{(0)}$ as a
starting value, and $y_{2}$ as data, obtain a sample from the posterior
distribution of $A_{1}$, $E_{1}$, $A_{2}$ and $c$ via the likelihood of
equation~(\ref{eq4.7}), and 1,000 iterations of the MCMC run. Sample values of $A_{1}$
and $E_{1}$ from their posterior\vadjust{\goodbreak} distributions obtained in Phase I will
serve as the priors of $A_{1}$ and $E_{1}$ in Phase II. Since the parameters
$A_{1}$ and $E_{1}$ reappear in the likelihood of equation (\ref{eq4.7}) as the mean
of $y_{2}$, Phase II of the MCMC run captures the effect of $y_{2}$ on these
parameters. The effect of $y_{1}$ was captured in Phase I.

\textit{Phase} III. Repeat Phase I and Phase
II $m$ times using new starting values of $E_{1}$ and $c$ to produce
a~sample of size $m$ from the posterior distribution of~$A_{1}$, $E_{1}$,
$A_{2}$ and $c$, with $y_{1}$ and $y_{2}$ as the data.

The MCMC exercise described above is routine, but computer intensive and
entails 12 steps, six in each phase, and this too for a highly curtailed
version of the Lattice QCD equations. The details of how this is done could
be interesting, because they involve some discretization of the simulated
posterior distributions, and working with individual sampled values
reminiscent of that done in particle Kal\-man filtering (cf. Gordon, Salmond and
Smith, \citeyear{GorSal}). Thus, we label our approach as {\textit{Particle
MCMC}}. More details are given in Landon (\citeyear{Lan07}), and the method illustrated in
the~\hyperref[app]{Appendix}. The software can be downloaded at
\href{http://www.gwu.edu/\textasciitilde stat/irra/Lattice\_QCD.htm}{http://www.gwu.edu/\textasciitilde stat/}
\href{http://www.gwu.edu/\textasciitilde stat/irra/Lattice\_QCD.htm}{irra/Lattice\_QCD.htm}.

\subsection{A Caveat of the Proposed Scheme}\label{sec4.3}

The caveat mentioned here stems from the features that $c$ has been fixed,
and that the MCMC runs are centered around fixed values of $y_{1}$ and $
y_{2} $. To see why, recall that our parsimonious version of the
Lattice QCD equations [see equation (\ref{eq3.4})] is based on those $t_{i}$'s for which the
exponential terms vanish; however, the $t_{i}$'s are determined by a fixed
value of $c$. Thus, any change in the value of $c$ will bring about a change
in the values of $t_{i}$, and, as a~consequence, the Lattice QCD equations
will also have to be different. This would be tantamount to obtaining new
values of the $y_{i}$'s. However, all the likelihoods in the MCMC runs are
based on fixed values of the $y_{i}$'s; see equations~(\ref{eq4.6}) and~(\ref{eq4.7}). But
a change in the value of $c$ is inevitable, because in Phase II of the MCMC
run one iterates around sampled values from the posterior distribution
of $c$, so that the initial $c^{(0)}$ systematically gets replaced by~$c^{(1)}$,
$c^{(2)},\ldots,c^{(1\mbox{,}000)}$.

A way to overcome this caveat is to recognize that for any
$c^{(i)}>c^{(0)}$, $i=1,2,\ldots,$ the exponential terms mentioned above will continue to
vanish, so that any specified values of $y_{i}$ will continue to
satisfy the
right-hand side of equation (\ref{eq4.2}).

\begin{figure*}[t!]
\centering
\begin{tabular}{@{}c@{\quad}c@{}}

\includegraphics{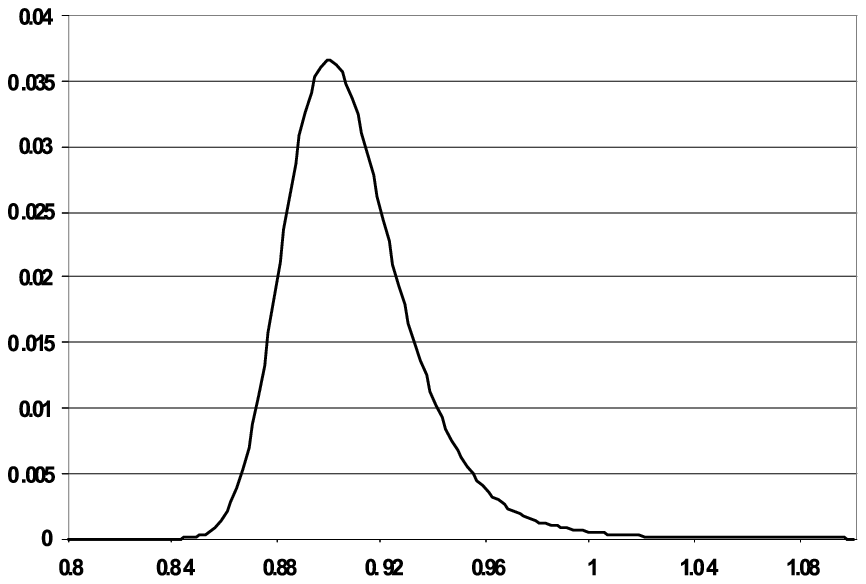}
 & \includegraphics{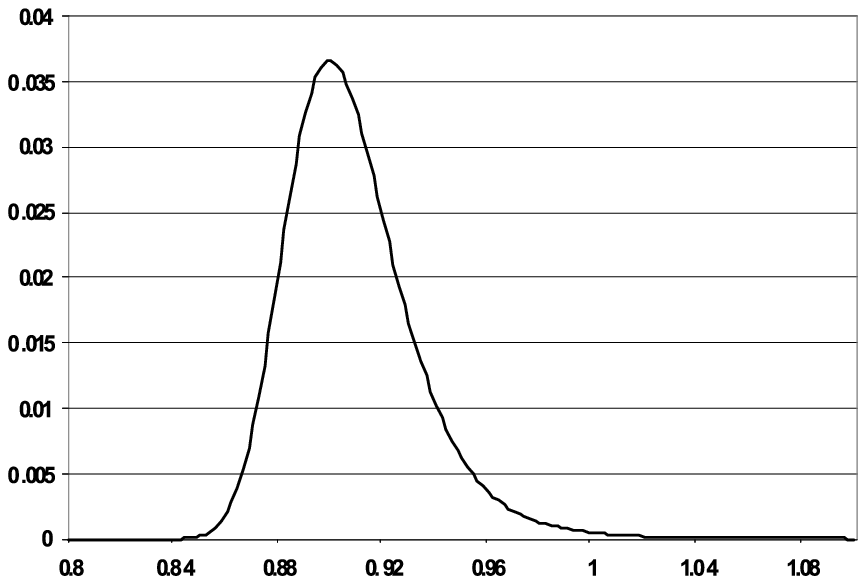}\\
\footnotesize{(a)} & \footnotesize{(b)}
\end{tabular}
\caption{Posterior distribution of $E_{1}$.
(\textup{a}) Posterior of $E_1$ based on $y_{12}$.
(\textup{b}) Posterior of $E_1$ based on $y_{12}$ and $y_{6}$.}\label{fig5.1}\vspace*{-3pt}
\end{figure*}

A strategy to ensure that the successively generated values of\vadjust{\goodbreak}
$c^{(i)}$, $i=1,2,\ldots,$ will tend to be greater than $c^{(0)}$ is to pick small
values of $c^{(0)}$ for each of the~$m$ iterations of Phase III of the MCMC
algorithm. During the course of the MCMC runs, should one encounter a
generated value of $c^{(i)}$ that is smaller than $c^{(0)}$, then one should
\textit{discard} the so-generated value $c^{(i)}$, and generate another
value of~$c^{(i)}$. Hopefully, the number of discarded $c^{(i)}$'s will not
be excessive, but if they are, then the starting value $c^{(0)}$ should be
decreased, and new values of~$t_{1}$ and $t_{2}$ obtained. This of course
would be tantamount to obtaining new values of $y_{1}$ and $y_{2}$ as
well.

\section{Proof of Principle: Validation Against Data}\label{sec5}

We first validate the accuracy of our approach against simulated data. For
this, we choose $A_{1}=0.8$, $A_{2}=0.6$, $A_{3}=0.4$, $A_{4}=0.2$, $%
A_{5}=0.1$, and $A_{i}=0$ for $i\geq6$. We also choose $E_{1}=0.9$ and
$c=0.5$. Using these values in equation (\ref{eq3.3}), we compute $G(t)$, for
$t=1,2,\ldots,12$; these are shown in column 3 of Table \ref{tab1}. Since
$Y_{t}=G(t|\cdot)+\varepsilon_{t}$, with $\varepsilon_{t}\sim N(0,\sigma
_{t}^{2})$ [see equation (\ref{eq4.2})], we generate $y_{1},\ldots,y_{12}$,
assuming the $\varepsilon_{t}$'s are independent, with $\sigma
_{t}=0.001\times G(t)\times t$; these are shown in column 4 of Table~\ref{tab1}. We
next identify those $t$'s for which the leading exponential terms vanish.
These happen to be $t_{1}$ at $t=12$, $t_{2}$ at $t=6$, and $t_{3}$ at
$t=4$; see column 2 of Table~\ref{tab1}. Our aim is to invoke the methods of
Section \ref{sec4}
on the entries of Table~\ref{tab1}, to see if the constants specified above
can be
returned. With the above in place, Phases I, II and III of the MCMC run were
made arbitrarily choosing the hyperparameters $\alpha=\beta=\eta
=\lambda
=\omega=1$, and $m=1\mbox{,}000$.

%
\begin{table}
\caption{Simulated data for validating approach}\label{tab1}
\begin{tabular*}{\columnwidth}{@{\extracolsep{\fill}}lccc@{}}
\hline
\multicolumn{1}{@{}l}{\textbf{Time} $\bolds{t}$} & \multicolumn{1}{c}{\textbf{Index} $\bolds{t_{i}}$} & \multicolumn{1}{c}{$\bolds{G(t)}$} &
\multicolumn{1}{c@{}}{$\bolds{y_{i}}$}\\
\hline
\phantom{0}1 & & 0.54874373 & 0.54900146\\
\phantom{0}2 & & 0.17764687 & 0.17756522\\
\phantom{0}3 & & 0.06387622 & 0.06373037\\
\phantom{0}4 & \multicolumn{1}{c}{$t_3$} & 0.02422158 & 0.02414992\\
\phantom{0}5 & & 0.00945326 & 0.00952723\\
\phantom{0}6 & \multicolumn{1}{c}{$t_2$} & 0.00375071 & 0.00377058\\
\phantom{0}7 & & 0.00151265 & 0.00151498\\
\phantom{0}8 & & 0.00060552 & 0.00061147\\
\phantom{0}9 & & 0.00024486 & 0.00024698\\
10 & & 0.00009923 & 0.00009821\\
11 & & 0.00004026 & 0.00004007\\
12 & \multicolumn{1}{c}{$t_1$} & 0.00001635 & 0.00001625\\
\hline
\end{tabular*}
\vspace*{-3pt}
\end{table}


\begin{figure*}[t!]
\centering
\begin{tabular}{@{}cc@{}}

\includegraphics{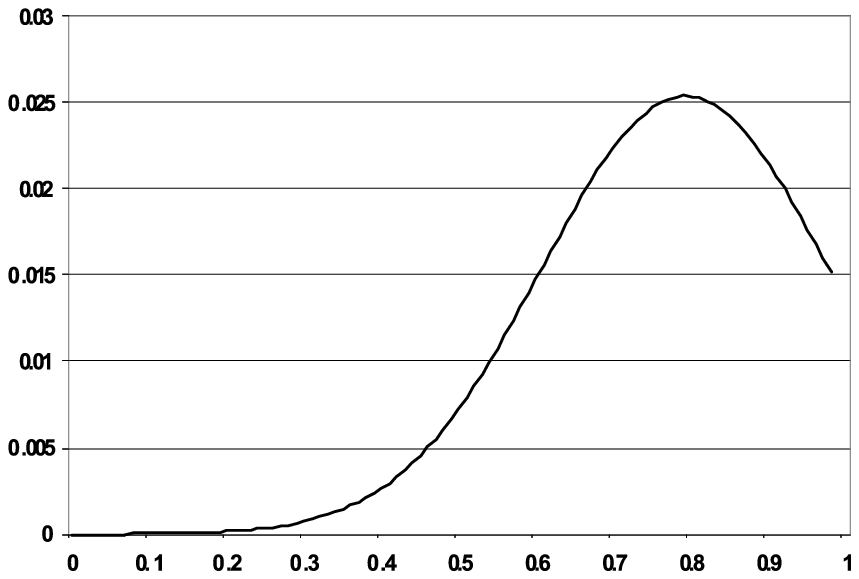}
 & \includegraphics{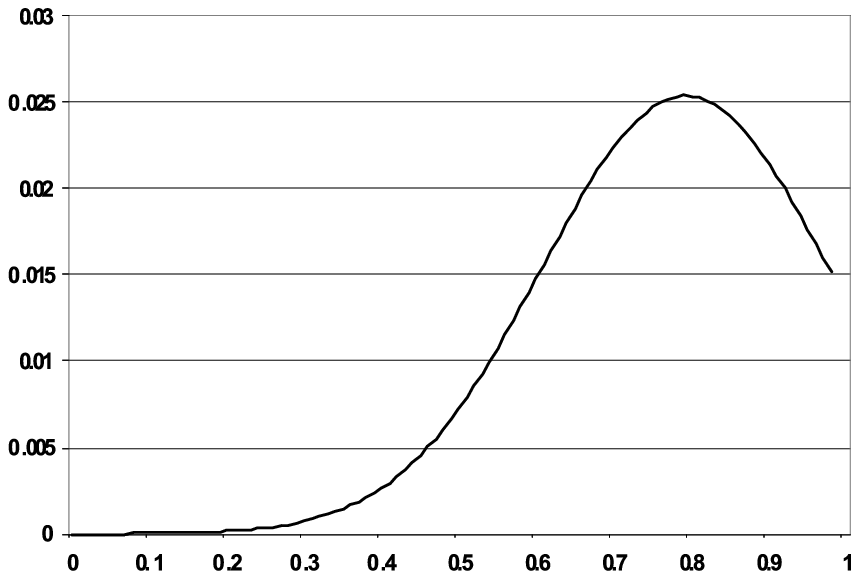}\\
\footnotesize{(a)} & \footnotesize{(b)}
\end{tabular}
\caption{Posterior distribution of $A_{1}$.
(\textup{a}) Posterior of $A_1$ based on $y_{12}$.
(\textup{b}) Posterior of $A_1$ based on $y_{12}$ and $y_{6}$.}\label{fig5.2}
\end{figure*}

\subsection{Results Based on Simulated Data}\label{sec5.1}

Figure \ref{fig5.1}(a) and (b) shows the posterior distributions of $E_{1}$ based
on $y_{12}$, and on $y_{12}$ and $y_{6}$, respectively. Recall that $y_{12}$
corresponds to $t_{1}$, and $y_{6}$ corresponds to $t_{2}$. Note that the
posterior distribution of Figure \ref{fig5.1}(a) becomes the prior
distribution for
the construction of the posterior distribution of Figure \ref{fig5.1}(b).
Both the
distributions of Figure \ref{fig5.1} indicate a modal value of 0.9, suggesting a
tendency to converge to the true value of $E_{1}$. Furthermore, the
difference between the two distributions is not very great, suggesting
that $%
y_{6}$ may not be contributing much toward inference for $E_{1}$, beyond
that provided by $y_{12}$.

A similar feature is revealed by the posterior distributions of $A_{1}$,
shown in Figures \ref{fig5.2}(a) and (b). These distributions have a modal value
of 0.8, suggesting again a convergence to the true value of\vadjust{\goodbreak} $A_{1}$.

\begin{figure}[t!]

\includegraphics{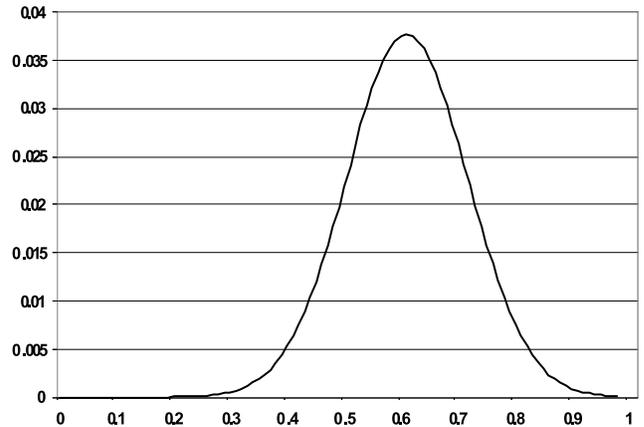}
  \caption{Posterior of $A_2$ based on $y_{12}$ and
  $y_6$.}\label{fig5.3}\vspace*{-6pt}
\end{figure}

\begin{figure}[t!]

\includegraphics{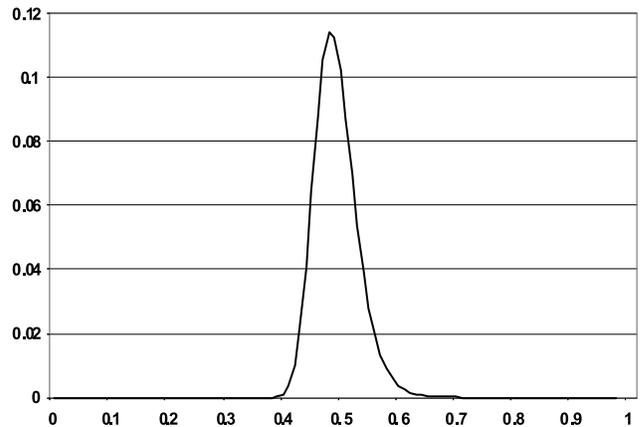}
  \caption{Posterior of $c$ based on $y_{12}$ and
  $y_6$.}\label{fig5.4}
  \vspace*{-3pt}
\end{figure}

Figures \ref{fig5.3} and \ref{fig5.4} show\vadjust{\goodbreak} the posterior distributions of $A_{2}$ and $c$,
based on $y_{12}$ and $y_{6}$. Their modal values of 0.6 and 0.5 suggest
convergence of the posteriors to their true values. Thus, based on this
simulation exercise, we may claim that, despite an arbitrary choice
of\vadjust{\goodbreak}
hyperparameters, the proposed MCMC procedure is able to show recovery
of the input values of $A_{1}$, $E_{1}$, $A_{2}$ and $c$ to a meaningful
degree of accuracy.

\subsubsection{Sensitivity of posteriors to priors}

In this section we explore the sensitivity of the posterior
distributions of
$A_{1}$, $E_{1}$, $A_{2}$ and $c$ when the hyperparameters of their
prior
distributions vary. We also explore the effect of using a thick-tailed prior
distribution for $E_{1}$, in particular, a Pareto distribution, instead of
the gamma distribution used before.

\begin{figure*}
\includegraphics{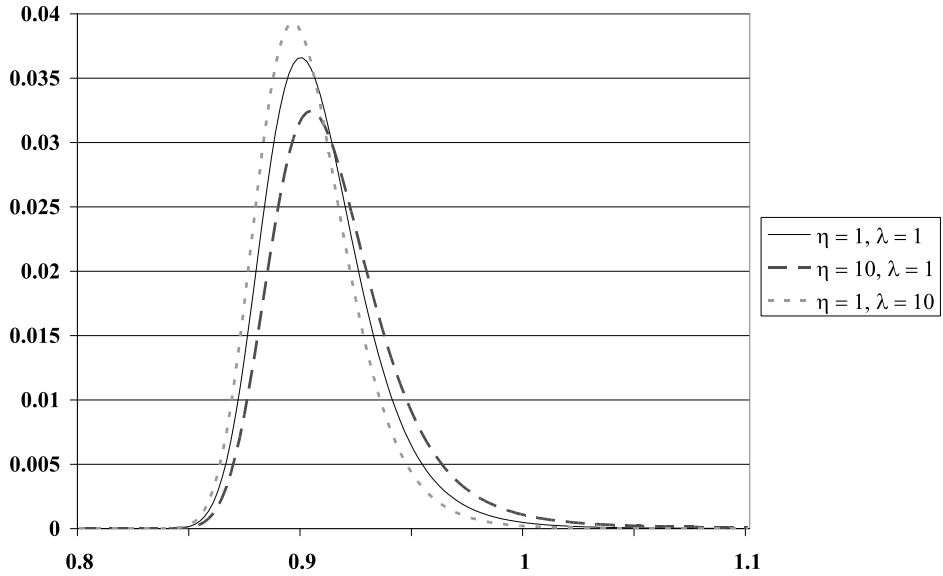}
\caption{Posterior distribution of $E_{1}$ with different values of $\eta$
and $\lambda$.}\label{fig5.5}
\vspace*{-6pt}
\end{figure*}

Figure \ref{fig5.5} shows the posterior distributions of $E_{1}$ for different values
of the scale $\lambda$ and shape $\eta$ parameters of its gamma prior.
Verify that the posterior distributions get centered around its true value
of~0.9 even when the prior mean is as large as 10. The values of the chosen
hyperparameters are indicated in the legend accompanying Figure \ref{fig5.5}.

\begin{figure*}
\includegraphics{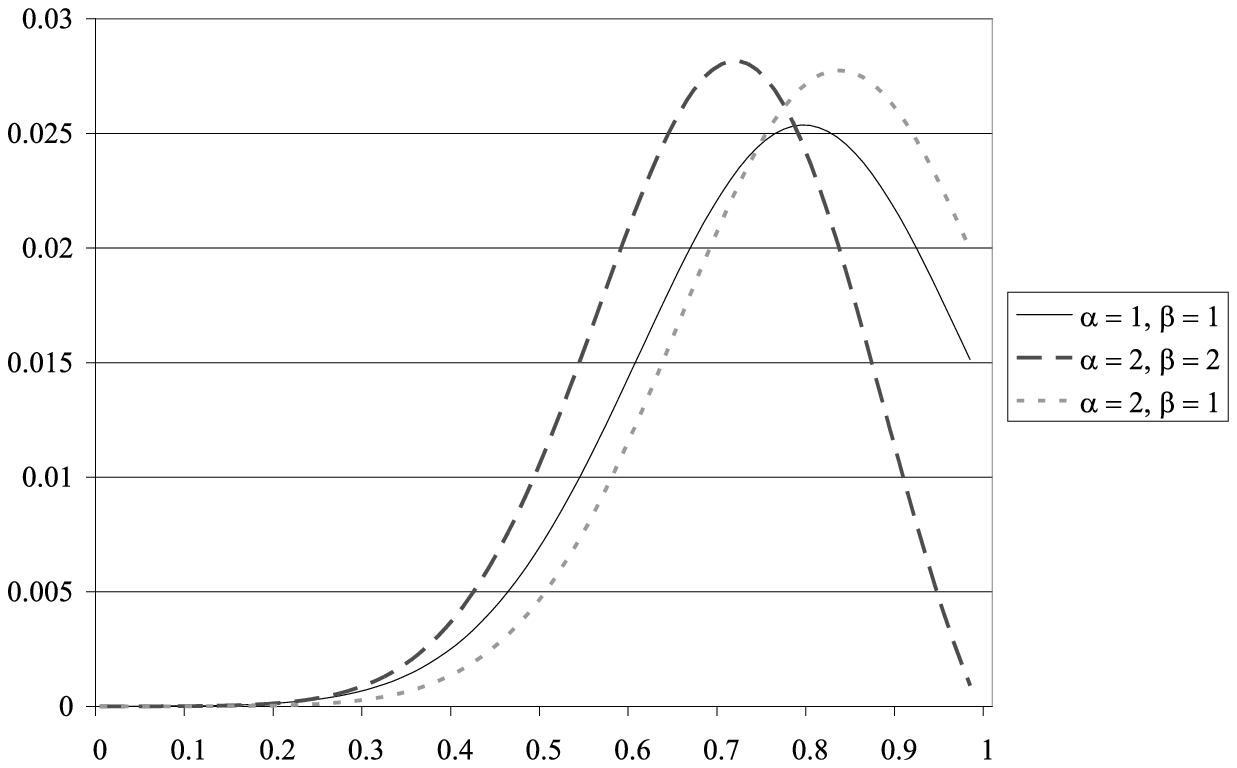}
  \caption{Posterior distribution of $A_{1}$ with different values of $\protect\alpha$ and $\protect\beta$.}\label{fig5.6}
\end{figure*}

\begin{figure*}
\includegraphics{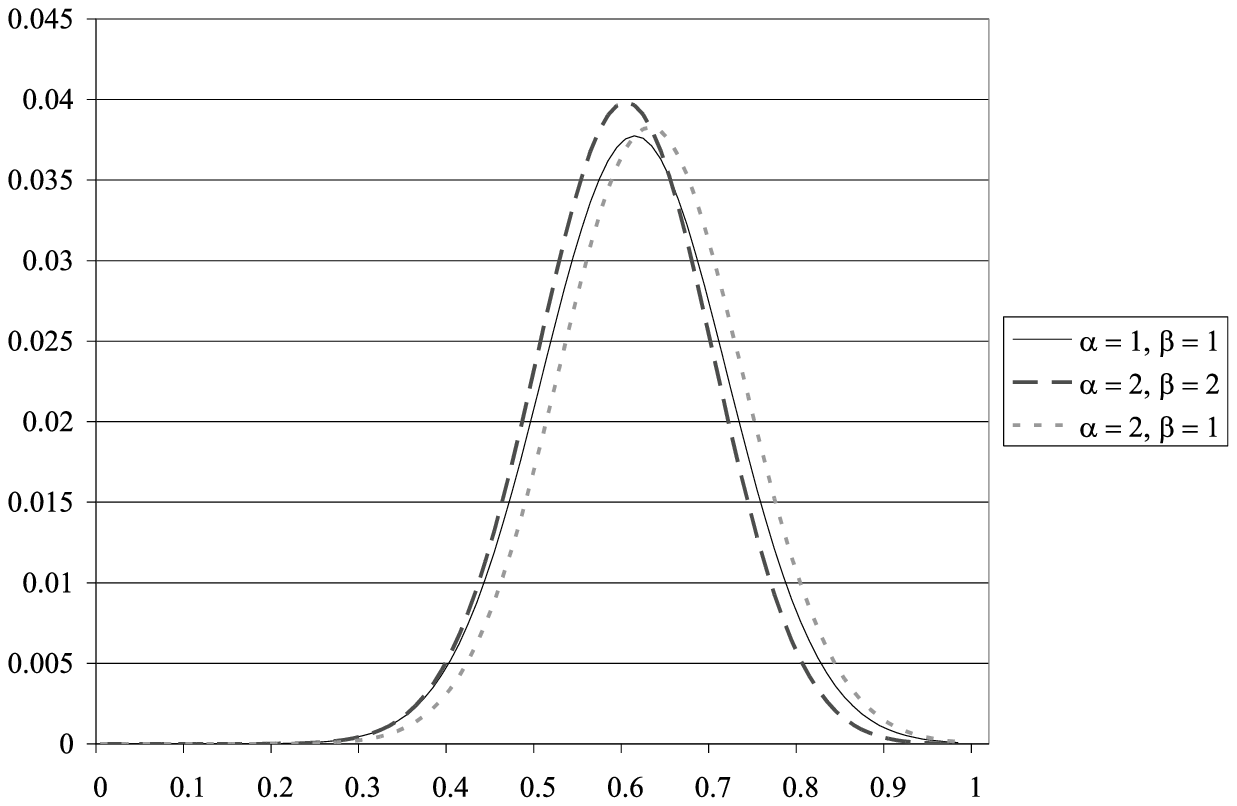}
  \caption{Posterior distribution of $A_{2}$ with different values of $\alpha$
and $\beta$.}\label{fig5.7}
\end{figure*}

In Figures \ref{fig5.6} and \ref{fig5.7} we show the posterior distributions of $A_{1}$
and $A_{2}$ for different values of the hyperparameters $\alpha$ and $\beta$;
see the legend accompanying these figures. Whereas the posterior
distribution of $A_{2}$ appears to be very robust against the various
choices for its prior distributions, the posterior distribution of $A_{1}$
shows some sensitivity---albeit minor---to the choice of its priors. These
priors are centered at (in the case of $A_{2}$) and around (in the case
of $A_{1}$) their true values of 0.6 and 0.8, respectively.

Since the prior on $c$ is a uniform on $(0,\omega/E_{1})$, changing the
value of $\omega$ would simply change the range of values that $c$ can
take. It will not change the shape of the posterior distribution of $c$.
Finally,\vadjust{\goodbreak} a~use of the Pareto as a prior for $E_{1}$ results in a posterior
distribution that looks much like that of Figure~\ref{fig5.1} produced by a gamma
prior. This result---not illustrated here---is true irrespective of the
choice of the hyperparameters of the Pareto prior. Indeed, the Pareto prior
for $E_{1}$ indicates a higher degree of robustness of its resulting
posterior as compared to the gamma prior.

Overall, it seems to be the case that the proposed procedure is robust to
the choice of priors, and that the resulting posteriors converge to their
correct values no matter the choice of\vadjust{\goodbreak} priors.

\subsection{Validation Against Physics Code Data}\label{sec5.2}

In this section we validate our approach using data pertaining to a pion
that has been generated by a~physics based code. These data are given in
Table~\ref{tab2} and parallel those of Table~\ref{tab1}, save for the fact that the data
run from $t=2$ to $t=13$, and that $G(t), t=\allowbreak 2,\ldots,13$, is not known.
However, the~$Y_{t}$'s and their associated errors are provided by the code,
the errors being a proxy for $\sigma_{t}^{2}$. The choice of~$t_{1},t_{2}$
and~$t_{3}$ is based on the following consideration. By
default,~$t_{1}$ has to be the largest $t$ for which the data are availab\-le;
thus, in our case $t_{1}$ corresponds to $t=13$. At~$t_{1}$~all the exponential terms
in equation (\ref{eq3.3}) vanish. At~$t_{2}$ we need to have the terms starting with
$e^{-2ct}$ vanish; this means that $t_{2}\approx t_{1}/2$, which in our case
would be 7. Similarly, $t_{3}\approx t_{1}/3$, which is 4, and so on.

\begin{figure*}[t]
\centering
\begin{tabular}{@{}cc@{}}

\includegraphics{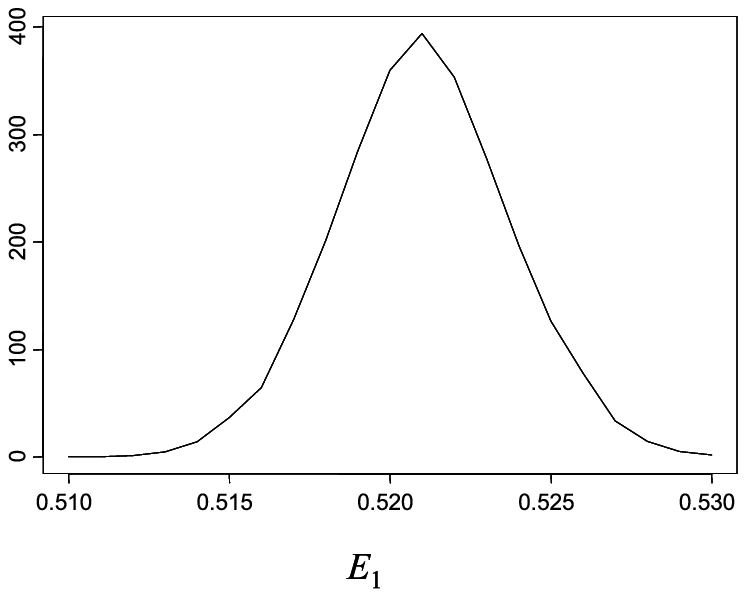}
 & \includegraphics{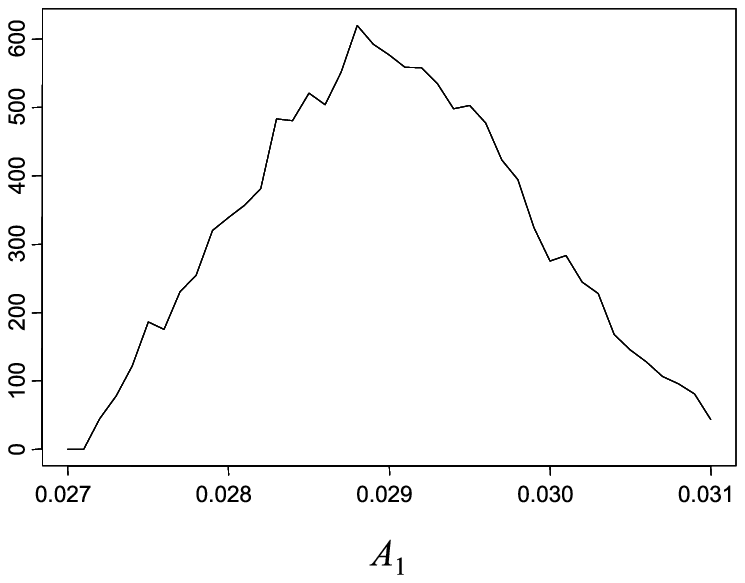}\\
\footnotesize{(a)} & \footnotesize{(b)}\\

\includegraphics{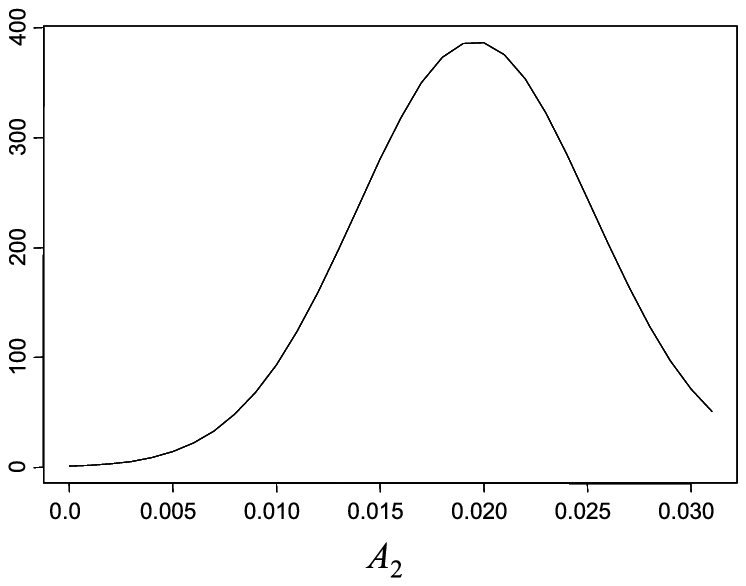}
 & \includegraphics{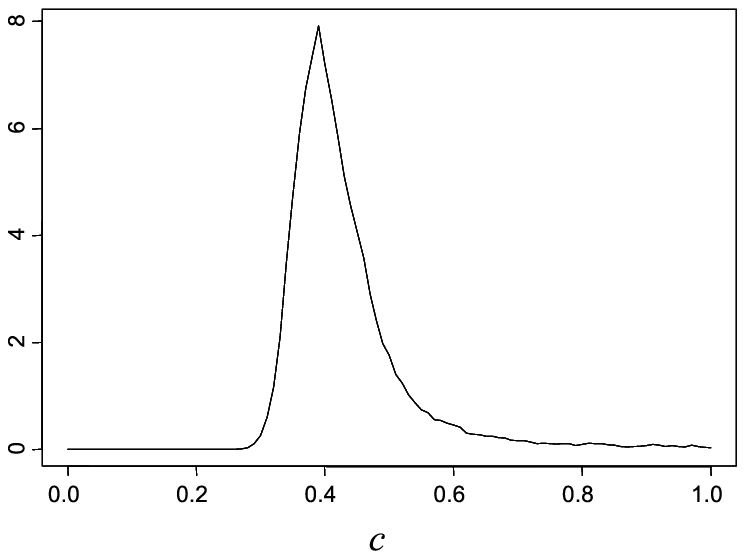}\\
\footnotesize{(c)} & \footnotesize{(d)}
\end{tabular}
  \caption{Posterior distributions of $A_{1}$, $E_{1}$, $A_{2}$ and $c$ based
on $y_{13}$ and $y_{7}$.
(\textup{a}) Final posterior of $E_1$.
(\textup{b}) Final posterior of~$A_1$.
(\textup{c}) Final posterior of $A_2$.
(\textup{d}) Final posterior of $c$.}\label{fig5.8}
\end{figure*}

\begin{table}
\caption{Physics code based data on a pion}
\label{tab2}
\begin{tabular*}{\columnwidth}{@{\extracolsep{\fill}}lcd{1.10}d{1.9}d{1.9}@{}}
\hline
\multicolumn{1}{@{}l}{$\bolds{t}$} & \multicolumn{1}{c}{\textbf{Index} $\bolds{t_i}$} & \multicolumn{1}{c}{$\bolds{Y_t}$} &
\multicolumn{2}{c@{}}{\textbf{Errors}}\\
\hline
\phantom{0}2 & & 0.043865236 & 0.00013635 & 0.00014836\\
\phantom{0}3 & & 0.009347211 & 0.00008205 & 0.000089027\\
\phantom{0}4 & \multicolumn{1}{c}{$t_3$} & 0.00406969 & 0.000051302 & 0.000057832\\
\phantom{0}5 & & 0.002187666 & 0.000031545 & 0.000034867\\
\phantom{0}6 & & 0.001252858 & 0.000018805 & 0.000018559\\
\phantom{0}7 & \multicolumn{1}{c}{$t_2$} & 0.000735911 & 0.000011131 & 0.00001124\\
\phantom{0}8 & & 0.0004358 & 6.6393\mathrm{E-}06 & 6.8252\mathrm{E-}06\\
\phantom{0}9 & & 0.00025829 & 0.000004049 & 4.3108\mathrm{E-}06\\
10 & & 0.000153161 & 2.4808\mathrm{E-}06 & 2.6302\mathrm{E-}06\\
11 & & 9.1412\mathrm{E-}05 & 1.5264\mathrm{E-}06 & 0.000001683\\
12 & & 5.552\mathrm{E-}05 & 9.5081\mathrm{E-}07 & 1.0741\mathrm{E-}06\\
13 & \multicolumn{1}{c}{$t_1$} & 3.54336\mathrm{E-}05 & 6.3079\mathrm{E-}07 & 7.0522\mathrm{E-}07\\
\hline
\end{tabular*}
\end{table}

In Figure \ref{fig5.8} $(a)$, $(b)$, $(c)$ and $(d)$, we show the posterior distributions
of $A_{1}$, $E_{1}$, $A_{2}$ and $c$, respectively, based on $y_{13}$
and $y_{7}$. The modes of these posterior distributions suggest the values of
0.52, 0.029, 0.02 and 0.4, for $E_{1}$, $A_{1}$, $A_{2}$ and $c$,
respectively. The values of $A_{1}$ and $E_{1}$ given above are in good
agreement with the values obtained by a physics based simulation code. Since
the physics based codes are unable to obtain good estimates of $A_{2}$
and $E_{2}$ (equivalently,~$c$), the results on $A_{2}$ and $c$ obtained by us
constitute a contribution toward the solution of an underlying scientific
problem.

Based on this exercise, plus others that are given~in Landon (\citeyear{Lan07}), our
conclusion therefore is that the proposed approach is successfully validated
against both simulated data as well as the physics code generated data. The
exercises in Landon (\citeyear{Lan07}) pertain to the quark masses of 4 photons and
5 pions.

\section{Extending the Approach}\label{sec6}

The approach outlined in Sections \ref{sec3} and \ref{sec4} has some limitations. The purpose
of this section is to prescribe strategies for overcoming these. By
far, the
most noteworthy limitation is that the model of equation (\ref{eq4.1}) restricts
attention to a consideration of the parameters $A_{1}$, $E_{1}$,
$A_{2}$ and
$c$, whereas the Lattice QCD equations have an infinite number of $A_{i}$'s
and $E_{i}$'s. The second concern pertains to the fact that in Section \ref{sec5},
data associated with the $t$'s intermediate to $t_{1}$ and $t_{2}$ are not
used in the MCMC algorithm. The proposed approach therefore does not exploit
all the available data $y_{t}$. Finally, there is a question of
assuming a
constant spacing $c$ of the~$E_{i}$'s. What is the effect of unequally
spaced~$E_{i}$'s on inference? Recall that the role played by $c$ is
important. First, it imparts parsimony by eliminating all the $E_{i}$'s save
for $E_{1}$. Second, it gives birth to the telescopic series which is
central to our approach. It turns\vadjust{\goodbreak} out that the effect of $c$ is
transitionary (it is a~nuisance parameter) and that inferences about~$E_{2}$,
$A_{3}$, $E_{3}$, $A_{4},\ldots,$ are possible if we exploit a result
observed in Section \ref{sec5}.

\subsection{Inferences for $E_{2}, A_{3}$ and Beyond}\label{sec6.1}

Our ability to extend the approach of Sections~\ref{sec3} and~\ref{sec4} to the case of
$E_{2}$, $A_{3}$, $E_{3}$, $A_{4},\ldots,$ is driven by the feature noticed in
Section \ref{sec5.1}, that $y_{6}$ does not contribute much toward the
inferences for
$A_{1}$ and~$E_{1}$, beyond that provided by $y_{12}$. Thus, the effect
of~$y_{4}$, which corresponds to $t_{3}$ of Table \ref{tab1}, will be less so,
making it
possible for us to do the following:

Rewrite equation (\ref{eq3.1}) as%
%
\begin{equation}\label{eq6.1}
G(t|\cdot)-A_{1}e^{-E_{1}t}=\sum_{n=2}^{\infty}A_{n}e^{-E_{n}t},
\end{equation}
and let $G^{\ast}(t|\cdot)=G(t|\cdot)-\widehat
{A}_{1}e^{-\widehat{E%
}_{1}t}$, where $\widehat{A}_{1}$ and $\widehat{E}_{1}$ are the modes
(means) of the posterior distributions of\vadjust{\goodbreak} $A_{1}$ and $E_{1}$ obtained via
the likes of Figures \ref{fig5.1}(b) and~\ref{fig5.2}(b). Thus,
\[
G^{\ast}(t|\cdot)\approx\sum_{n=2}^{\infty}A_{n}e^{-E_{n}t},
\]
and setting $A_{n}^{\ast}=A_{n+1}$ and $E_{n}^{\ast}=E_{n+1}$, for $%
n=2,\allowbreak 3,\ldots,$ we have%
\begin{equation}\label{eq6.2}
G^{\ast}(t|\cdot)\approx\sum_{n=1}^{\infty}A_{n}^{\ast
}e^{-E_{n}^{\ast}t}.
\end{equation}
The right-hand side of equation (\ref{eq6.2}) parallels the right-hand side of
equation (\ref{eq3.1}), save for the fact that $A_{n}^{\ast}$ and $E_{n}^{\ast}$
replace $A_{n}$ and $E_{n}$. The material of Sections \ref{sec3} and \ref{sec4} now applies,
but with the caveat that since equation (\ref{eq6.2}) is an approximation, whereas
equation~(\ref{eq3.1}) is exact, the variance of the error terms associated
with the
former should be larger than those associated with the latter.

The posterior distributions of $E_{1}^{\ast}$ and $A_{2}^{\ast}$ will be
the posterior distributions of $E_{2}$ and $A_{3}$. The role of $c$ as a
nuisance parameter is now apparent. The posterior distribution of $%
A_{1}^{\ast}$ will serve as a revised posterior distribution of $A_{2}$.
Indeed, for the MCMC runs associated with the treatment of equation~(\ref{eq6.2}), we
may sample from the posterior distribution of $A_{2}$ to generate the
posterior distribution of $A_{1}^{\ast}$.

We may continue in the above vein to estimate~$E_{3}$ and $A_{4}$ by
defining $G^{\ast\ast}(t|\cdot)=G^{\ast}(t|\cdot)-\widehat
{A}%
_{1}^{\ast}e^{-\widehat{E}_{1}^{\ast}t}$, whe\-re $\widehat
{A}_{1}^{\ast}$
and $\widehat{E}_{1}^{\ast}$ are the modes of the posterior distributions
of $A_{1}^{\ast}$ and $E_{1}^{\ast}$, respectively, and similarly
with $%
(E_{4},A_{5})$, $(E_{5},A_{6})$, and so on.

\subsection{Using Additional $Y_{t}$'s}\label{sec6.2}

For enhanced inferences about the parameters $A_{1}$ and $E_{1}$ we may want
to use all values of $Y_{t}$ inter\-mediate to those associated with the
labels $t_{1}$ and~$t_{2}$ of Tables \ref{tab1} and \ref{tab2} and, similarly, with
the $Y_{t}$'s intermediate to the ones associated with the labels~$t_{2}$ and~$t_{3}$,
and so on. What makes this possible is the fact that $t_{1}$ is the
\textit{largest} value of $t$ for which $(A_{2},E_{2})$, $(A_{3},E_{3}),
\ldots,$ gets annihilated, whereas $t_{2}$ is the lar\-gest value of $t$ at
which $(A_{3},E_{3}),(A_{4},E_{4}),\ldots,$ gets annihilated, and so on.
Thus, values of $t$ intermediate to $t_{1}$ and $t_{2}$ will continue to
annihilate $(A_{2},E_{2})$, $(A_{3},E_{3}),\ldots,$ and those
intermediate to $t_{2}$ and $t_{3}$ will annihilate $(A_{3},E_{3}),(A_{4},E_{4}),\ldots,$
and so on.

Let $y_{11},y_{12},\ldots,y_{15}$ denote the $Y_{t}$'s intermediate to
those associated with the labels $t_{1}$ and $t_{2}$. Then, to incorporate
the effect of $y_{11},\ldots,y_{15}$ for enhanced inference about $A_{1}$
and $E_{1}$, the iterative scheme described in Phase I of Section \ref{sec4.2} will
have to be cycled five more times, each cycle involving a use of the
$y_{1j}$, $j=1,\ldots,5$, before proceeding to Phase II, wherein the effect of $
y_{6} $ (of Table \ref{tab1}) and $y_{7}$ (of Table \ref{tab2}) comes into play and,
similarly, with $y_{21}$, the single value intermediate to that associated
with the labels $t_{2}$ and $t_{3}$.

\section{Summary and Conclusions}\label{sec7}

In this paper we have proposed and developed a~statistical approach for
addressing a much discus\-sed problem in particle physics. Indeed, a problem
that has spawned several Nobel prizes in Physics. The essence of the problem
boils down to estimating a~large (conceptually infinite) number of unknown
parameters based on a finite number of nonlinear equations. Statisticians
refer to such problems as large $p$---small $n$. Each equation in our
problem comprises of the sum of several exponential
functions.\looseness=1

Previous approaches for addressing this problem have been physics based---such as perturbation
me\-thods---and statistics based---such as chi-squared\break
goodness of fit, and Empirical Bayes. Physicists have found such approaches
unsatisfactory, and have cal\-led for a use of proper Bayesian approaches,
thus this paper.

The Bayesian approach proposed by us has been facilitated by the fact that
by introducing a latent parameter, the architecture of the nonlinear
equations reveals an attractive pattern. This pattern boils down to our
consideration of a truncated telescopic series of equations, each equation
being the sum of a finite number of exponential functions. Similar sets of
equations also arise in other arenas of science, as mentioned in Section
\ref{sec3.1}. The nonlinear nature of the equations mandates that our proposed
approach---which entails stylized proper priors---be implemented by a
particle style Markov chain Monte Carlo (MCMC) approach. Such a procedure
turns out to be computationally very intensive---about one million
iterations for making inference about three parameters.

The proposed procedure, when invoked on simulated data, is able to reproduce
the input parameters. This is one way to claim the validity of our approach.
The procedure, when invoked on some real data pertaining to the quark masses
of protons and pions, is also able to produce results that are in agreement
with the results produced using alternate physics based methods. However,
the physics based methods are able to obtain only partial results. By
contrast, our approach can produce estimates of as many parameters as is
desired---but there is no way to validate these against alternate
approaches or actual numbers, because these are not available.\looseness=1

Future work in this arena will entail enhancements to gain computational
efficiencies and the choice of proper priors that are motivated by a
consideration of the underlying physics. This means that an undertaking such
as this will call for insights and skills that go beyond mathematics,
statistics and computing. Some appreciation of the underlying physics is
necessary for, among other things, interest and inspiration! A referee of
this paper has made the interesting suggestion of considering
``reference priors.''  This we have been
unable to do because, for the parameters in question, such priors are not
readily available.

\section*{Acknowledgments}\vspace*{6pt}
The authors would like to thank Professor Ali Eskandarian of GWU for
introducing us to this topic and for orchestrating our involvement in
it. We
want to acknowledge (the late) Professor Cornelius Benn\-hold, also of GWU,
for contributing to our learning of the physics part of the problem, for
keeping us honest and for providing us with the data to validate our
approaches. The several helpful comments of the referees and the Editor,
Professor David Madigan, are gratefully acknowledged. This research is
supported by The Army Research Office Grant W911NF-09-1-0039 and by the
National Science Foundation Grant DMS-09-15156 with The George Washington
University. The work of Frank Lee is supported in part by U.S.
Department of
Energy under Grant DE-FG02-95ER-40907.

\begin{appendix}\label{app}

\vspace*{3pt}\section*{Appendix}\vspace*{3pt}

Schemata of the 3 Phase MCMC algorithm, which can be downloaded at the
following:
\href{http://www.gwu.edu/\textasciitilde stat/irra/Lattice\_QCD.htm}{http://www.gwu.}
\href{http://www.gwu.edu/\textasciitilde stat/irra/Lattice\_QCD.htm}{edu/\textasciitilde stat/irra/Lattice\_QCD.htm}.
\vspace*{15pt}

\includegraphics{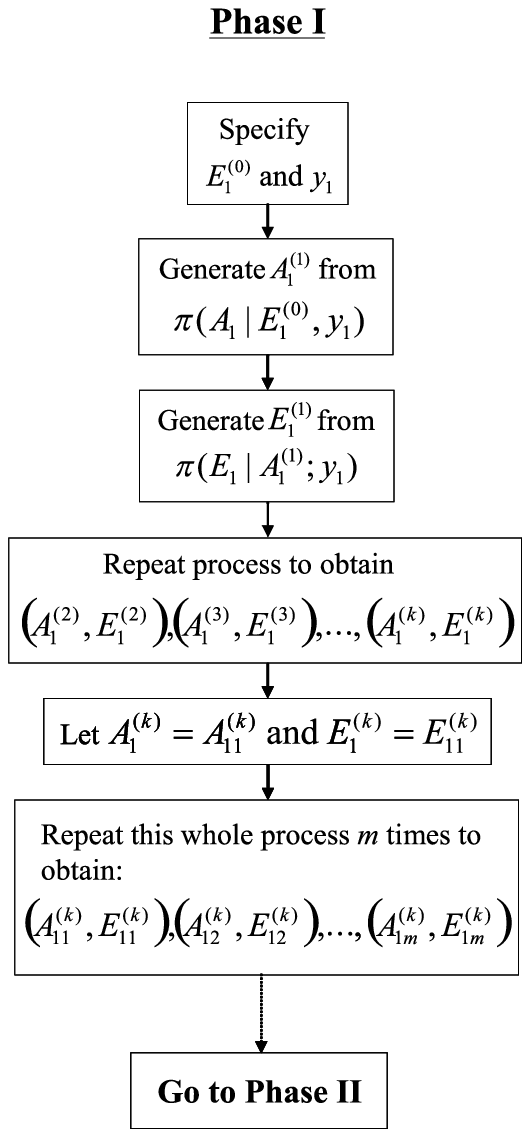}

\includegraphics{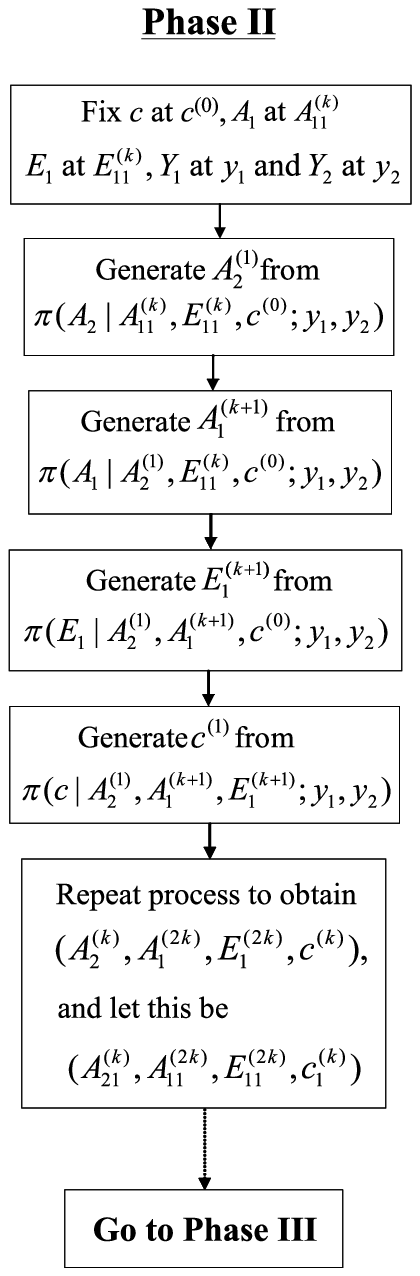}
\vspace*{12pt}

\includegraphics{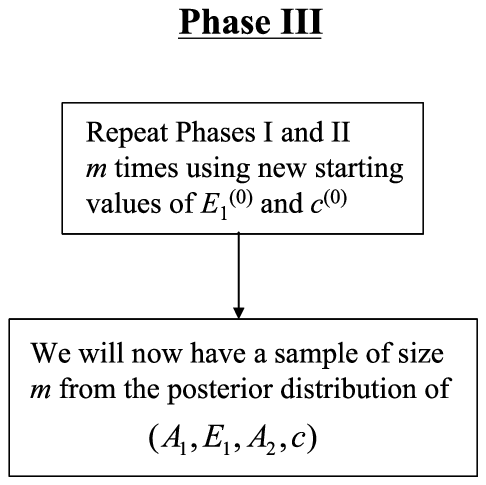}

\end{appendix}
%
%



\begin{thebibliography}{30}



\bibitem[\protect\citeauthoryear{Barlow and Proschan}{1975}]{BarPro75}
\begin{bbook}[mr]
\bauthor{\bsnm{Barlow},~\bfnm{Richard~E.}\binits{R.~E.}} \AND
  \bauthor{\bsnm{Proschan},~\bfnm{Frank}\binits{F.}}
(\byear{1975}).
\btitle{Statistical Theory of Reliability and Life Testing}.
\bpublisher{Holt}, \baddress{Rinehart and Winston, New York}.
\bid{mr={0438625}}
\end{bbook}
\endbibitem

\bibitem[\protect\citeauthoryear{Bernardo and Smith}{1994}]{BerSmi94}
\begin{bbook}[mr]
\bauthor{\bsnm{Bernardo},~\bfnm{Jose-M.}\binits{J.-M.}} \AND
  \bauthor{\bsnm{Smith},~\bfnm{Adrian F.~M.}\binits{A.~F.~M.}}
(\byear{1994}).
\btitle{Bayesian Theory}.
\bpublisher{Wiley}, \baddress{Chichester}.
\bid{doi={10.1002/9780470316870}, mr={1274699}}
\end{bbook}
\endbibitem
%




\bibitem[\protect\citeauthoryear{Bretthorst et~al.}{2005}]{Breetal05}
\begin{barticle}[auto:STB|2011-03-03|12:04:44]
\bauthor{\bsnm{Bretthorst},~\bfnm{G.~L.}\binits{G.~L.}},
  \bauthor{\bsnm{Hutton},~\bfnm{W.~C.}\binits{W.~C.}},
  \bauthor{\bsnm{Garbow},~\bfnm{J.~R.}\binits{J.~R.}} \AND
  \bauthor{\bsnm{Ackerman},~\bfnm{J.~J.~H.}\binits{J.~J.~H.}}
(\byear{2005}).
\btitle{Exponential parameter estimation (in NMR) using Bayesian probability
  theory}.
\bjournal{Concepts in Magnetic Resonance Part A}
\bvolume{27A}
\bpages{55--63}.
\end{barticle}
\endbibitem

\bibitem[\protect\citeauthoryear{Chen et~al.}{2004}]{Cheetal}
\begin{bmisc}[auto:STB|2011-03-03|12:04:44]
\bauthor{\bsnm{Chen},~\bfnm{Y.}\binits{Y.}},
  \bauthor{\bsnm{Draper},~\bfnm{T.}\binits{T.}},
  \bauthor{\bsnm{Dong},~\bfnm{S.~J.}\binits{S.~J.}},
  \bauthor{\bsnm{Horvath},~\bfnm{I.}\binits{I.}},
  \bauthor{\bsnm{Lee},~\bfnm{F.~X.}\binits{F.~X.}},
  \bauthor{\bsnm{Liu},~\bfnm{K.~F.}\binits{K.~F.}},
  \bauthor{\bsnm{Mathur},~\bfnm{N.}\binits{N.}},
  \bauthor{\bsnm{Srinivasan},~\bfnm{C.}\binits{C.}},
  \bauthor{\bsnm{Tamhankar},~\bfnm{S.}\binits{S.}} \AND
  \bauthor{\bsnm{Zhang},~\bfnm{J.~B.}\binits{J.~B.}}
(\byear{2004}).
\bhowpublished{The sequential empirical Bayes method: An adaptive
  constrained-curve fitting algorithm for lattice QCD.'' \textit{Phys. Rev.
  D}. Available at \url{http://arxiv.org/pdf/hep-lat/0405001}}.
\end{bmisc}
\endbibitem

\bibitem[\protect\citeauthoryear{Dyson and Isenberg}{1971}]{DysIse71}
\begin{barticle}[pbm]
\bauthor{\bsnm{Dyson},~\bfnm{R.~D.}\binits{R.~D.}} \AND
  \bauthor{\bsnm{Isenberg},~\bfnm{I.}\binits{I.}}
(\byear{1971}).
\btitle{Analysis of exponential curves by a method of moments, with special
  attention to sedimentation equilibrium and fluorescence decay}.
\bjournal{Biochemistry}
\bvolume{10}
\bpages{3233--3241}.
\bid{issn={0006-2960}, pmid={4107520}}
\end{barticle}
\endbibitem


\bibitem[\protect\citeauthoryear{Dzierba, Meyer and Swanson}{2000}]{AmericanScientist00}
\begin{bmisc}[auto:STB|2011-03-03|12:04:44]
\bauthor{\bsnm{Dzierba},~\bfnm{A.}\binits{A.}},
  \bauthor{\bsnm{Meyer},~\bfnm{C.}\binits{C.}} \AND
  \bauthor{\bsnm{Swanson},~\bfnm{E.}\binits{E.}}
(\byear{2000}).
\bhowpublished{The search for QCD exotics. \textit{American Scientist} \textbf{88}(5) 406--416}.
\end{bmisc}
\endbibitem

\bibitem[\protect\citeauthoryear{Fiebig}{2002}]{Fie}
\begin{bmisc}[auto:STB|2011-03-03|12:04:44]
\bauthor{\bsnm{Fiebig},~\bfnm{H.~R.}\binits{H.~R.}}
(\byear{2002}).
\bhowpublished{Spectral density analysis of time correlation functions in
  lattice QCD using the maximum entropy method. \textit{Phys. Rev. D} \textbf{65} 094512}.
\end{bmisc}
\endbibitem

\bibitem[\protect\citeauthoryear{Fleming}{2005}]{Fl05}
\begin{bmisc}[auto:STB|2011-03-03|12:04:44]
\bauthor{\bsnm{Fleming},~\bfnm{George T.}\binits{G.~T.}}
(\byear{2005}).
\bhowpublished{What can lattice QCD theorists learn from nmr spectrocopists?
Technical Report, Jefferson Labs., Newport News, VA}.
\end{bmisc}
\endbibitem


\bibitem[\protect\citeauthoryear{Foss}{1969}]{Fos69}
\begin{barticle}[auto:STB|2011-03-03|12:04:44]
\bauthor{\bsnm{Foss},~\bfnm{S.~D.}\binits{S.~D.}}
(\byear{1969}).
\btitle{A method for obtaining initial estimates of the parameters in
  exponential curve fitting}.
\bjournal{Biometrics}
\bvolume{25}
\bpages{580--584}.
\end{barticle}
\endbibitem

\bibitem[\protect\citeauthoryear{Giurc{\u{a}}neanu, T{\u{a}}bu{\c{s}} and
  Astola}{2005}]{GiuTabAst05}
\begin{barticle}[mr]
\bauthor{\bsnm{Giurc{\u{a}}neanu},~\bfnm{Ciprian~Doru}\binits{C.~D.}},
  \bauthor{\bsnm{T{\u{a}}bu{\c{s}}},~\bfnm{Ioan}\binits{I.}} \AND
  \bauthor{\bsnm{Astola},~\bfnm{Jaakko}\binits{J.}}
(\byear{2005}).
\btitle{Clustering time series gene expression data based on
  sum-of-exponentials fitting}.
\bjournal{EURASIP J. Appl. Signal Process.}
\bvolume{8}
\bpages{1159--1173}.
\bid{issn={1110-8657}, mr={2168617}}
\end{barticle}
\endbibitem

\bibitem[\protect\citeauthoryear{Glass and de Garreta}{1967}]{GlaGer67}
\begin{barticle}[auto:STB|2011-03-03|12:04:44]
\bauthor{\bsnm{Glass},~\bfnm{H.~I.}\binits{H.~I.}} \AND
  \bauthor{\bsnm{de Garreta},~\bfnm{A.~C.}\binits{A.~C.}}
(\byear{1967}).
\btitle{Quantitative analysis of exponential curve fitting for biological
  applications}.
\bjournal{Physics in Medicine and Biology}
\bvolume{12}
\bpages{379--388}.
\end{barticle}
\endbibitem

\bibitem[\protect\citeauthoryear{Gordon, Salmond and Smith}{1993}]{GorSal}
\begin{barticle}[auto:STB|2011-03-03|12:04:44]
\bauthor{\bsnm{Gordon},~\bfnm{N.~J.}\binits{N.~J.}},
  \bauthor{\bsnm{Salmond},~\bfnm{D.~J.}\binits{D.~J.}} \AND
  \bauthor{\bsnm{Smith},~\bfnm{A.~F.~M.}\binits{A.~F.~M.}}
(\byear{1993}).
\btitle{Novel approach to nonlinear/non-Gaussian Bayesian state estimation}.
\bjournal{IEE Proceedings F}
\bvolume{140}
\bpages{107--113}.
\end{barticle}
\endbibitem

\bibitem[\protect\citeauthoryear{Griffiths}{1987}]{Gri87}
\begin{bbook}[auto:STB|2011-03-03|12:04:44]
\bauthor{\bsnm{Griffiths},~\bfnm{D.}\binits{D.}}
(\byear{1987}).
\btitle{Introduction to Elementary Particles}.
\bpublisher{Wiley}, \baddress{New York}.
\end{bbook}
\endbibitem

\bibitem[\protect\citeauthoryear{Hildebrand}{1956}]{Hil56}
\begin{bbook}[mr]
\bauthor{\bsnm{Hildebrand},~\bfnm{F.~B.}\binits{F.~B.}}
(\byear{1956}).
\btitle{Introduction to Numerical Analysis}.
\bpublisher{McGraw-Hill}, \baddress{New York}.
\bid{mr={0075670}}
\end{bbook}
\endbibitem

\bibitem[\protect\citeauthoryear{Landon}{2007}]{Lan07}
\begin{bmisc}[mr]
\bauthor{\bsnm{Landon},~\bfnm{Joshua}\binits{J.}}
(\byear{2007}).
\bhowpublished{A problem in particle physics and its {B}ayesian analysis.
Ph.D. thesis, George Washington Univ., Washington, DC.}
\bid{mr={2710570}}
\end{bmisc}
\endbibitem

\bibitem[\protect\citeauthoryear{Lepage et~al.}{2002}]{Lepetal02}
\begin{barticle}[auto:STB|2011-03-03|12:04:44]
\bauthor{\bsnm{Lepage},~\bfnm{G.~P.}\binits{G.~P.}},
  \bauthor{\bsnm{Clark},~\bfnm{B.}\binits{B.}},
  \bauthor{\bsnm{Davies},~\bfnm{T.~H.}\binits{T.~H.}},
  \bauthor{\bsnm{Hornbostel},~\bfnm{K.}\binits{K.}},
  \bauthor{\bsnm{Mackenzie},~\bfnm{P.~B.}\binits{P.~B.}},
  \bauthor{\bsnm{Morningstar},~\bfnm{C.}\binits{C.}} \AND
  \bauthor{\bsnm{Trottier},~\bfnm{H.}\binits{H.}}
(\byear{2002}).
\btitle{Constrained curve fitting}.
\bjournal{Nuclear Physics B Proceedings Supplements}
\bvolume{106}
\bpages{12--20}.
\end{barticle}
\endbibitem\

\bibitem[\protect\citeauthoryear{Morningstar}{2002}]{Mor02}
\begin{barticle}[auto:STB|2011-03-03|12:04:44]
\bauthor{\bsnm{Morningstar},~\bfnm{C.}\binits{C.}}
(\byear{2002}).
\btitle{Bayesian curve fitting for lattice gauge theorists}.
\bjournal{Nuclear Physics B Proceedings Supplements}
\bvolume{109}
\bpages{185--191}.
\end{barticle}
\endbibitem

\bibitem[\protect\citeauthoryear{Nakahara, Asakawa and
  Hatsuda}{1999}]{NakAsaHat99}
\begin{barticle}[auto:STB|2011-03-03|12:04:44]
\bauthor{\bsnm{Nakahara},~\bfnm{Y.}\binits{Y.}},
  \bauthor{\bsnm{Asakawa},~\bfnm{M.}\binits{M.}} \AND
  \bauthor{\bsnm{Hatsuda},~\bfnm{T.}\binits{T.}}
(\byear{1999}).
\btitle{Hadronic spectral functions in lattice QCD}.
\bjournal{Phys. Rev. D}
\bvolume{60}
\bpages{091503}.
\end{barticle}
\endbibitem

\bibitem[\protect\citeauthoryear{Pagels}{1982}]{Pag82}
\begin{bbook}[auto:STB|2011-03-03|12:04:44]
\bauthor{\bsnm{Pagels},~\bfnm{H.~R.}\binits{H.~R.}}
(\byear{1982}).
\btitle{The Cosmic Code: Quantum Physics As the Language of Nature.}
\bpublisher{Simon and Schuster}, \baddress{New York}.
\end{bbook}
\endbibitem

\bibitem[\protect\citeauthoryear{Paluszny et~al.}{2008/09}]{Paletal08}
\begin{barticle}[mr]
\bauthor{\bsnm{Paluszny},~\bfnm{Marco}\binits{M.}},
  \bauthor{\bsnm{Mart{\'{\i}}n-Landrove},~\bfnm{Miguel}\binits{M.}},
  \bauthor{\bsnm{Figueroa},~\bfnm{Giovanni}\binits{G.}} \AND
  \bauthor{\bsnm{Torres},~\bfnm{Wuilian}\binits{W.}}
(\byear{2008/09}).
\btitle{Boosting the inverse interpolation problem by a sum of decaying
  exponentials using an algebraic approach}.
\bjournal{Electron. Trans. Numer. Anal.}
\bvolume{34}
\bpages{163--169}.
\bid{issn={1068-9613}, mr={2597808}}
\bptnote{check year}%
\end{barticle}
\endbibitem

\bibitem[\protect\citeauthoryear{Perl}{1960}]{Per60}
\begin{barticle}[auto:STB|2011-03-03|12:04:44]
\bauthor{\bsnm{Perl},~\bfnm{W.}\binits{W.}}
(\byear{1960}).
\btitle{A method for curve-fitting by exponential functions}.
\bjournal{The International Journal of Applied Radiation and Isotopes}
\bvolume{8}
\bpages{211--222}.
\end{barticle}
\endbibitem


\bibitem[\protect\citeauthoryear{Riordan and Zajc}{2006}]{ScientificAmerican06}
\begin{bmisc}[auto:STB|2011-03-03|12:04:44]
\bauthor{\bsnm{Riordan},~\bfnm{M.}\binits{M.}} \AND
\bauthor{\bsnm{Zajc},~\bfnm{W. A.}\binits{W. A.}}
(\byear{2006}).
\bhowpublished{The first few microseconds. \textit{Scientific American} \textbf{294} 34--41}.
\end{bmisc}
\endbibitem

\bibitem[\protect\citeauthoryear{Robertson}{1957}]{Rob57}
\begin{barticle}[auto:STB|2011-03-03|12:04:44]
\bauthor{\bsnm{Robertson},~\bfnm{J.~S.}\binits{J.~S.}}
(\byear{1957}).
\btitle{Theory and use of tracers in determining transfer rates in biological
  systems}.
\bjournal{Physiological Reviews}
\bvolume{37}
\bpages{133--157}.
\end{barticle}
\endbibitem

\bibitem[\protect\citeauthoryear{Rubinow}{1975}]{Rub75}
\begin{bbook}[auto:STB|2011-03-03|12:04:44]
\bauthor{\bsnm{Rubinow},~\bfnm{S.~I.}\binits{S.~I.}}
(\byear{1975}).
\btitle{Introduction to Mathematical Biology}.
\bpublisher{Wiley}, \baddress{New York}.
\end{bbook}
\endbibitem

\bibitem[\protect\citeauthoryear{Sanchez-Gasca and Chow}{1999}]{SanCho99}
\begin{barticle}[auto:STB|2011-03-03|12:04:44]
\bauthor{\bsnm{Sanchez-Gasca},~\bfnm{J.~J.}\binits{J.~J.}} \AND
  \bauthor{\bsnm{Chow},~\bfnm{J.~H.}\binits{J.~H.}}
(\byear{1999}).
\btitle{Performance comparison of three identification methods for the analysis
  of electromagnetic oscillations}.
\bjournal{IEEE Transactions on Power Systems}
\bvolume{14}
\bpages{995--1002}.
\end{barticle}
\endbibitem

\bibitem[\protect\citeauthoryear{Singpurwalla}{2006}]{Sin06}
\begin{bbook}[mr]
\bauthor{\bsnm{Singpurwalla},~\bfnm{Nozer~D.}\binits{N.~D.}}
(\byear{2006}).
\btitle{Reliability and Risk: A Baye\-sian Perspective}.
\bpublisher{Wiley}, \baddress{Chichester}.
\bid{doi={10.1002/9780470060346}, mr={2265917}}
\end{bbook}
\endbibitem

\bibitem[\protect\citeauthoryear{Smith and Morales}{1944}]{SmiMor44}
\begin{barticle}[auto:STB|2011-03-03|12:04:44]
\bauthor{\bsnm{Smith},~\bfnm{R.~E.}\binits{R.~E.}} \AND
  \bauthor{\bsnm{Morales},~\bfnm{M.~F.}\binits{M.~F.}}
(\byear{1944}).
\btitle{On the theory of blood-tissue exchanges: II. Applications}.
\bjournal{Bull. Math. Biol.}
\bvolume{6}
\bpages{133--139}.
\end{barticle}
\endbibitem

\bibitem[\protect\citeauthoryear{Van~Liew}{1967}]{Van67}
\begin{barticle}[auto:STB|2011-03-03|12:04:44]
\bauthor{\bsnm{Van~Liew},~\bfnm{H.~D.}\binits{H.~D.}}
(\byear{1967}).
\btitle{Method of exponential peeling}.
\bjournal{J.~Theoret. Biol.}
\bvolume{16}
\bpages{43--53}.
\end{barticle}
\endbibitem

\bibitem[\protect\citeauthoryear{Wilczek}{2005}]{Wil05}
\begin{barticle}[mr]
\bauthor{\bsnm{Wilczek},~\bfnm{Frank~A.}\binits{F.~A.}}
(\byear{2005}).
\btitle{Asymptotic freedom: From paradox to paradigm}.
\bjournal{Internat. J. Modern Phys. A}
\bvolume{20}
\bpages{5753--5777}.
\bid{doi={10.1142/S0217751X05029022}, issn={0217-751X}, mr={2189131}}
\bptnote{check year}%
\end{barticle}
\endbibitem


\bibitem[\protect\citeauthoryear{Yam}{1993}]{ScientificAmerican93}
\begin{bmisc}[auto:STB|2011-03-03|12:04:44]
\bauthor{\bsnm{Yam},~\bfnm{P.}\binits{P.}}
(\byear{1993}).
\bhowpublished{QED for QCD.  \textit{Scientific American} \textbf{269}  23--24}.
\end{bmisc}
\endbibitem

\vspace*{-3pt}
\end{thebibliography}
\end{document}